**Localized corrosion resistance of ultrafine-grained austenitic steel 321 obtained by Rotary Swaging: The role of annealing conditions**

A.V. Nokhrin[(*)], M.K. Chegurov, N.V. Melekhin, D.A. Zotov, V.I. Kopylov, A.N. Sysoev, K.A. Rubtsova

*Lobachevsky State University of Nizhny Novgorod, 603022, Russia, Nizhny Novgorod, Gagarin ave., 23*

e-mail: nokhrin@nifti.unn.ru

**Abstract**

The object of this study was austenitic steel 321, which has an ultrafine-grained (UFG) structure formed by Rotary Swaging (RS). RS at room temperature leads to the formation of a two-phase macroscopically heterogeneous and, at the same time, a microscopically homogeneous UFG structure with an increased content of martensite. The UFG steel has a high ultimate tensile strength (UTS = 1580 MPa) and a low elongation to failure (δ ~4%). It has been shown that secondary TiC nanoparticles form in the annealing temperature range of 450–550°C, the reverse transformation of martensite into austenite occurs in the temperature range of 550–650°C, and recrystallization begins at 700°C. Electrochemical studies of the resistance of the UFG steel to general corrosion, intergranular corrosion (IGC), and pitting corrosion (PC) have been carried out. It has been shown that the UFG steel after RS has a reduced resistance to IGC. It has been found that after RS, there is an uneven distribution of microhardness and corrosion resistance in the cross-section of UFG steel rods. The surface of steel rods after RS has a reduced hardness and corrosion resistance compared to the central part of the rods. It has been shown that in the temperature range of 450–550°C, there is a slight increase in the microhardness, a slight decrease in the ductility, and a significant decrease in the IGC and PC resistance of the UFG steel. Recrystallization annealing at 700–800°C leads to an increased PC resistance and IGC resistance. Annealing conditions have been determined that ensure a good combination of strength, ductility and corrosion resistance of the UFG steel 321. It was demonstrated that to explain the non-monotonic variation in the localized corrosion rate of UFG steel 321 during annealing, it is necessary to consider not only the well-known effects of martensite and δ-ferrite, but also the precipitation of TiC particles and the changes in the free energy of non-equilibrium grain boundaries.

(*) Corresponding author (Aleksey Nokhrin, Dr.Sc., Head of Laboratory, e-mail: nokhrin@nifti.unn.ru, https://orcid.org/0000-0002-0328-1923)

## 1. Introduction

The petrochemical, mechanical, and nuclear engineering fields all make extensive use of coarse-grained (CG) Fe-Cr-Ni austenitic stainless steels. The high corrosion resistance of austenitic Fe-Cr-Ni steels enables their use in the manufacture of various critical nuclear engineering components that operate under the influence of aggressive media of different types [1, 2]. The low strength at room temperature (RT) of industrial austenitic steels is one of their main disadvantages. In this regard, the urgent task is to increase the strength of austenitic steels without significantly reducing their corrosion resistance. An effective way to increase the strength of metallic materials is to form an ultrafine-grained (UFG) structure in them (see, for example, [3]).

Dynamic recrystallization during hot working is a common method for refining the microstructure of austenitic stainless steels [2, 4, 5]. This process results in the formation of a fine-grained austenitic microstructure, which enhances the strength of the materials. However, the ever-increasing demands for reliability require the use of steels with significantly higher strength.

In this article, we focus on studying the mechanical properties and the corrosion resistance of UFG metastable austenitic steel grade 321 produced by Rotary Swaging (RS).

As is known, during cold plastic deformation of metastable steels, martensite is formed [6]. The RS method is based on the fast working of the metal rod surface due to the impact of high-strength strikers on it [7-9]. This process causes adiabatic heating of the rod’s surface, which in turn suppresses the $\gamma \rightarrow \alpha$ phase transformation [10] and promotes the formation of a homogeneous UFG austenite structure in the surface layer of the material. High strain rates and the localization of plastic deformation also contribute to the formation of smaller martensite grains [11–13], which leads to the formation of a finer-grained austenite structure during subsequent heat treatment. In accordance with the Hall-Petch relationship, this allows for an additional increase in the strength of austenitic steels.

Over the past few years, detailed studies have been conducted on the structure of UFG metastable steel grade 321 obtained by the RS method [14–22]. Notably, studies [14, 17, 19, 22] provide a detailed investigation of the three primary stages of evolution in the phase composition, dislocation structure, and grain microstructure of austenitic steel 321 during cold RS. In [14, 17, 22], it was demonstrated that, in the steel rod after cold RS, a structure is formed consisting of cellular austenite, lamellar austenite, and $\alpha'$-martensite. The reverse transformation $\alpha' \rightarrow \gamma$ in the UFG steel 321 occurs at 520–920°C [18]. At 650–700°C, the processes of recrystallization and grain growth begin [14, 18], which provide increased ductility of steel.

The specifics of the reverse transformation of martensite into austenite during heating of UFG steel 321 obtained by the RS method are described in [18]. Several studies [14–17, 19–21] have examined the mechanical characteristics of UFG metastable steel 321 obtained by RS. The authors of [14, 15] successfully achieved an optimal balance of ultimate tensile strength and elongation-to-failure in UFG steel 321, while also improving its impact toughness, through the process of cold RS followed by heat treatment.

A key characteristic of materials processed by RS is the non-uniform microhardness distribution across the rod cross-section (see reviews [7, 8]). These reviews [7-9] highlight that the maximum values of microhardness can occur either at the rod surface or in its core. Consequently, RS is regarded as a highly effective technique for fabricating heterostructured and/or functionally graded metallic materials. The authors of [14, 15, 19-21] demonstrated that both microhardness distribution profiles can be observed in 321 steel rods, depending on the applied strain during RS. At low strain levels at RS, the maximum values of microhardness are observed at the rod surface. At high strains, the maximum microhardness values are observed both at the center of the cross-section and on the rod's surface, while a decrease in hardness occurs at a distance of ~1–2 mm from the surface [14, 15, 17, 21]. This suggests to consider rods made of steel 321 after high total strains by the RS method as materials with a macroscopically heterogeneous structure (see also reviews [22–25]).

Research of heterostructured metastable steels produced by RS is highly active (for example, [14–16, 20]). However, most studies focus on the mechanical properties of heterostructured UFG steels, while their corrosion resistance remain largely unexplored.

It is known that the martensite reduces the corrosion resistance of austenitic steel [2, 26]. Therefore, it is generally expected that increased martensite content in UFG steels may lead to a decrease in their corrosion resistance [2, 12, 27, 28]. However, some authors argue that the formation of the UFG structure does not affect the corrosion resistance of austenitic steel [29], or that its effect is unclear [30, 31]. Several studies have reported higher corrosion resistance for UFG austenitic steel and for austenitic steels with an increased content of strain-induced martensite [12, 32-36]. It was shown in [37] that subsequent heat treatment can enhance the corrosion resistance of UFG steel 304. It is also known that SPD can influence the corrosion resistance of austenitic steels by refinements of non-metallic inclusions [38], due to changes in the Cr concentration at the grain boundaries [40] or in the surface layer [12]. However, the corrosion resistance of heterostructured austenitic steel obtained by RS has not been previously investigated.

The goal of this research is to investigate the impact of annealing on the corrosion resistance of UFG austenitic steel grade 321 produced using the RS method. A distinctive aspect of this work is the use of the RS method without forced cooling, which leads to intensive heating of the steel rod. Most authors use the RS method with forced cooling of the steel rod, facilitating the formation of an increased content of martensite (up to 80% [14, 15, 17, 19–21]). As noted above, subsequent annealing is used to develop a homogeneous fine-grained structure, effectively treating the RS method as a form of cold plastic deformation. However, self-heating of the steel rod during RS reduces the intensity of martensite formation and can directly enable the formation of a UFG austenitic structure with increased strength.

The microstructure and mechanical properties of UFG steel 321 were discussed in [40], so we will not go into further detail on this topic. However, in section 3.1 of our article we provided information about the microstructure and mechanical properties of the UFG steel, which are necessary for the analysis of the results. The mechanical characteristics and corrosion resistance of CG steel 321 have been thoroughly investigated (for example, see [41]). Furthermore, numerous studies have focused on UFG steel 321 produced by SPD methods (for example, [36, 42–47]). This enables us to compare our results with those of other researchers, including the study results of UFG steel 321 produced by alternative deformation processing methods.

## 2. Methods

Commercial austenitic metastable steel grade 321 was the object of this study. The chemical composition of the steel is presented in Table 1. The steel rods were not subjected to preliminary quenching.

Table 1. Chemical composition of steel grade 321 (wt. %)

| Fe | C | Cr | Ni | Mn | Si | Ti | Cu | S | P |
|---|---|---|---|---|---|---|---|---|---|
| Balance | 0.06 | 18.3 | 10.2 | 0.61 | 0.23 | 0.49 | 0.22 | 0.003 | 0.001 |

The steel rods with an initial diameter of 20 mm were obtained using an RS5-4-21 HIP machine. RS was carried out up to diameter $D$ = 6 mm. The total strain was $e$ = 91%. No forced cooling of the rod surface was applied during swaging. The average strain rate was 0.6-1 $s^{-1}$. The workpiece rotated at a 60 rpm. The average strike rate was 1200 strikes per minute. The feed rate of the workpiece was 50 $mm \cdot min^{-1}$.

The structure of the steel was studied using a Jeol JSM-6490 scanning electron microscope (SEM) and a Leica IM DRM optical microscope (OM). The average grain and secondary particle sizes were determined using the intercept method. Samples for structural studies were prepared by mechanical grinding and polishing. GoodGrains 2.0 software was used to quantify the size and volume fraction of grains, secondary particles and inclusions. Electrolytic polishing was conducted in an alcoholic solution of $CrO_3 + H_3PO_4$ for 60 s at a voltage of 30 V and a current of 3 A. Electrolytic etching was performed in the same solution for 90 s at a voltage of 3 V and a current of 0.3 A.

The content of the martensite and δ-ferrite was determined by magnetic phase analysis using a MAG-51B laboratory magnetometer (Russia). When calculating the amount of the ferromagnetic phase, it was assumed that an increase in its amount leads to a linear increase in saturation magnetization ($I_S$). The samples were oriented along the rod axis. The error in determining the volume fraction of the ferromagnetic phase was ± 0.2%. The martensite content was calculated by subtracting the saturation magnetization of coarse-grained steel ($I_{S(CG)}$) from that of UFG steel ($I_{S(UFG)}$).

Additional studies of the phase composition of the steel were carried out using a Haoyuan DX-2700BH X-ray diffractometer (CuKα radiation). X-ray studies were conducted on metallographic sections. To study the coarse-grained sample, a traditional slit system (DS = 1°, SS = 1°, RS = 0 mm) was used, which makes it possible to analyze an area of ~10×5 mm$^2$. The analysis of the XRD results was performed using the databases PDF-2 (ICDD, release 2012), PDF-4 (ICDD, release 2014) and ICSD (2016). Qualitative phase analysis was performed using the DIFFRAC.EVA software.

The unit cell parameters of the detected phases were determined using the Rietveld method by modeling the experimental X-ray diffraction patterns in the DIFFRAC.TOPAS software. The initial (theoretical) crystallographic parameters for modeling were taken from the cif-files of the ICSD. The diffraction patterns of the coarse-grained samples were modeled assuming that α-Fe exhibits texture along the (110) direction, while γ-Fe exhibits texture along the (111) direction. For the deformed samples, the XRD patterns were modeled assuming that α-Fe has texture along the (110) direction, and γ-Fe has texture along both the (111) and (002) directions.

Microhardness (Hv) was measured using a Qness A60+ hardness tester with a load of 0.3 kg. Hardness measurements were conducted in the cross-section of the rod with a step of 0.2–0.3 mm from one edge of the rod to the other in such a way that the indentation line passed through the sample center. The average error in determining Hv was 50 MPa. Microhardness measurements were carried out on metallographic sections.

Compression and tensile tests were carried out at RT on a 2167 R-50 machine (Russia). The compression tests were carried out on samples with a diameter of 5 mm and a height of 10 mm. Cylindrical samples with a head zone diameter of 5 mm and a working part diameter of 3 mm were tension tested (Fig. 1b). The sample cutting scheme is shown in Fig. 1a, b. During the compression tests, the stress-strain curves were obtained, from which the yield compressive strength (YCS) of the stainless steel was determined. The average error of measuring the YCS value was ± 20 MPa. The conditional yield tensile strength (YTS), ultimate tensile strength (UTS), elongation to failure (El), and reduction of area at fracture (ψ) were determined during tensile testing. The average measurement error of the YTS and UTS values was ± 20 MPa, and the values of $\delta_5$ and ψ were 0.2% and 5%, respectively.

Corrosion tests were conducted using a potentiostat-galvanostat R-8 at RT in an aqueous solution of 10% $H_2SO_4$ with 0.0025 g/L KSCN in a 60 mL three-electrode glass cell. A silver-chloride reference electrode (EVL-1M4) and a platinum electrode, which served as the auxiliary electrode, were used in the testing. Flat samples measuring 2×6×15 mm were used. Their surfaces were subjected to mechanical grinding and polishing to achieve a roughness level of no more than 1 μm, followed by washing and degreasing. The samples were oriented along the axis of the rod. The sample cutting scheme is shown in Fig. 1c. For the local study of corrosion resistance of the surface layer, a protective mask of corrosion-resistant varnish was applied to the sample's surface, leaving an unprotected area of ~ 0.6 to 0.8 $mm^2$ that closely adjoins the sample's surface (in Fig. 1c this area is marked with #1). Local electrochemical studies of the central layer of the rod were conducted in a similar manner (area #2 in Fig. 1c). The area of active corrosion, which is unprotected by a corrosion-resistant varnish and exhibits a different color, was measured using the GoodGrains software. The results of local studies on the central and surface layers were compared with the findings from electrochemical studies conducted throughout the entire sample area (the area shaded in gray in Fig. 1d). In the past, we employed this method to assess the corrosion resistance of titanium rods obtained by RS [48].

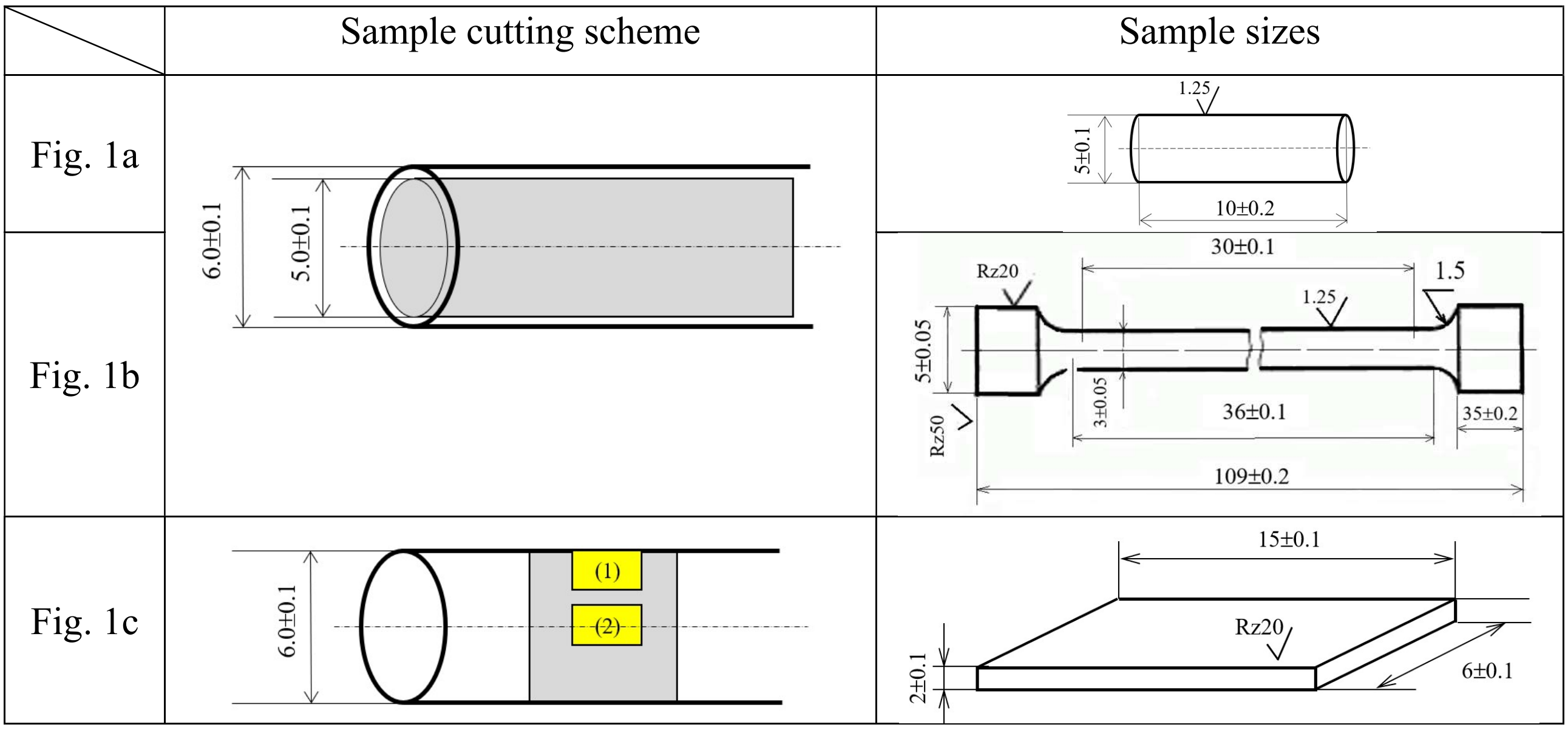

**Fig. 1**. Cutting schemes and sample sizes: cutting scheme for mechanical compression samples (a) and tensile test samples (b); (c) arrangement of active corrosion areas on the sample surface during electrochemical tests

As a result of electrochemical studies of the steel samples in an aqueous solution of 10% $H_2SO_4$ with 0.0025 g/L KSCN, several types of curves were obtained corresponding to different types of corrosion and corresponding test methods:

(i) Before conventional electrochemical tests, the samples were preliminarily exposed in the test cell in the corrosion solution for 1 h until a stable open-circuit potential was reached while simultaneously monitoring the potential-time curve $E(t)$. The $E(t)$ dependencies illustrated the electrochemical behavior of the steel samples at the initial moment of stationary tests in the full immersion mode.

(ii) The $i(E)$ curve of current on potential from –0.8 to 1.2 V (vs. Ag/AgCl) at a potential scanning rate of 0.5 mV/s and a recording rate of 10 points/s. The samples were preliminarily conditioned in the test cell for 1 h. The $i(E)$ curves were used primarily to determine the range of potentials corresponding to different electrochemical conditions: the hydrogen evolution region, the passivation region, the repassivation region, etc.

(iii) The Tafel method was used to determine the general corrosion rate. The Tafel test was carried out in the potential range from −1.0 to 0.4 V (vs. Ag/AgCl) with a scanning rate of 0.5 mV/s and a recording rate of 10 points/s. The steel samples were preliminarily conditioned in the test cell for 1 h. By analyzing the $\ln(i)$–$E$ curves, the corrosion current density ($i_{corr}$, $mA/cm^2$) and corrosion potential ($E_{corr}$, mV vs. Ag/AgCl) were determined. The experimental uncertainty in determining $i_{corr}$ was in the range of 0.1–0.15 $mA/cm^2$, while the error in $E_{corr}$ determination was ± 5 mV (vs. Ag/AgCl). The corrosion rate ($V_{corr}$) was calculated using the formula: $V_{corr} = 8.76 \cdot i_{corr} M/\rho F$, where $\rho$ is the steel density [$g/cm^3$], $M$ is the molar mass of iron [g/mol], $F$ = 96 500 C stands for the Faraday constant. Based on a joint analysis of the $E(t)$ curves and the $\ln(i)$–$E$ Tafel curves, the susceptibility of the samples to general corrosion was assessed.

(iv) The $i(t)$ curves of current on test time at constant potential values $E$ = const. Based on the analysis of the $i(E)$ curves, the potential values were chosen to be $E_1$ = 0.6 V and $E_2$ = 1.2 V (vs. Ag/AgCl). The value of potential $E_1$ corresponded to the passivation region of the steel samples, and the value of potential $E_2$ corresponded to the beginning of the fracture region of the protective film and the formation of pits. Based on the analysis of the $i(t)$ dependencies, the resistance of the UFG austenitic steel to pitting corrosion was assessed.

(v) The IGC resistance of the steels was investigated using an R-8 potentiostat-galvanostat, employing the double loop electrochemical potentiodynamic reactivation (DLEPR) method [49]. In this method, the sample is initially passivated by varying the potential at a specified rate, transitioning from the corrosion potential to the passivation potential. The potential is then altered in the opposite direction. During the test, reactivation of the sample occurs, accompanied by the breakdown of the passive film. It is generally assumed that the degradation of the passive film begins at the surface of the austenitic steel grain boundaries, which may be depleted in chromium or contain of the σ-phase particles, etc.

According to the Russian National Standard GOST 9.914-91, the increasing of the coefficient of $K$ to critical value $K_{IGC}$ = 0.11 means that the austenite steel 321L demonstrates an increased tendency to IGC. The sample under study was polarized at a potential of –0.55 V (vs. Ag/AgCl) for 2 min. The DLEPR voltage-current curve was obtained in the potential range of –0.55 to 1.2 V (vs. Ag/AgCl), with a scan rate of 3 mV/s. The IGC susceptibility was determined by the ratio of the areas under the reactivation ($S_1$) and activation ($S_2$) curves, expressed as $K = S_1/S_2$. In terms of its meaning, the coefficient $K$ is analogous to the Degree of Sensitization (DOS), which is defined as the ratio of the reactivation current ($I_r$) to the activation current ($I_a$). In accordance with GOST 9.914-91, an increase in the $K$ coefficient to the critical value of $K_{IGC}$ ~0.11 indicates that austenitic steel 321 exhibits an increased susceptibility to IGC. For each structural state, at least three identical samples were examined, each cut from different sections of the rod. For the same area of active corrosion, at least 3–5 measurements were performed. The average error in determining the coefficient $K$ was 0.03.

The surfaces of the specimens after the corrosion tests were examined using a Leica IM DRM microscope.

Annealing of the steel samples was carried out for 30 minutes in the EKPS-10 air furnace, on a steel substrate. The temperature during annealing was maintained within ± 5°C. The samples were allowed to cool in still air after the annealing process.

## 3. Results

### 3.1 Microstructure and mechanical properties

*Coarse-grained steel*

In the as-received state, the average austenite grain size of steel 321 is 20–30 μm (Fig. 2a, b). Recrystallization twins are observed within the austenite grains. Stainless steel 321 exhibits a homogenous coarse-grained microstructure with isolated inclusions of δ-ferrite, as well as inclusions of Ti(C,N) (Fig. 2a). Thin δ-ferrite bands traverse the austenite grains. Their average thickness is 10–20 μm, while their length varies from 100 μm to 0.5–1 mm. The optic and MFA study results show that the volume fraction of δ-ferrite is 5 ± 1%. The results of the EDS analysis demonstrate that δ-ferrite particles contain an increased concentration of Cr (up to 24%) and a reduced content of Ni (less than 4%). The size of the Ti(C,N) individual inclusions varies from 5 to 20 μm; the inclusions form agglomerates up to 100–120 μm long, oriented along the rod axis.

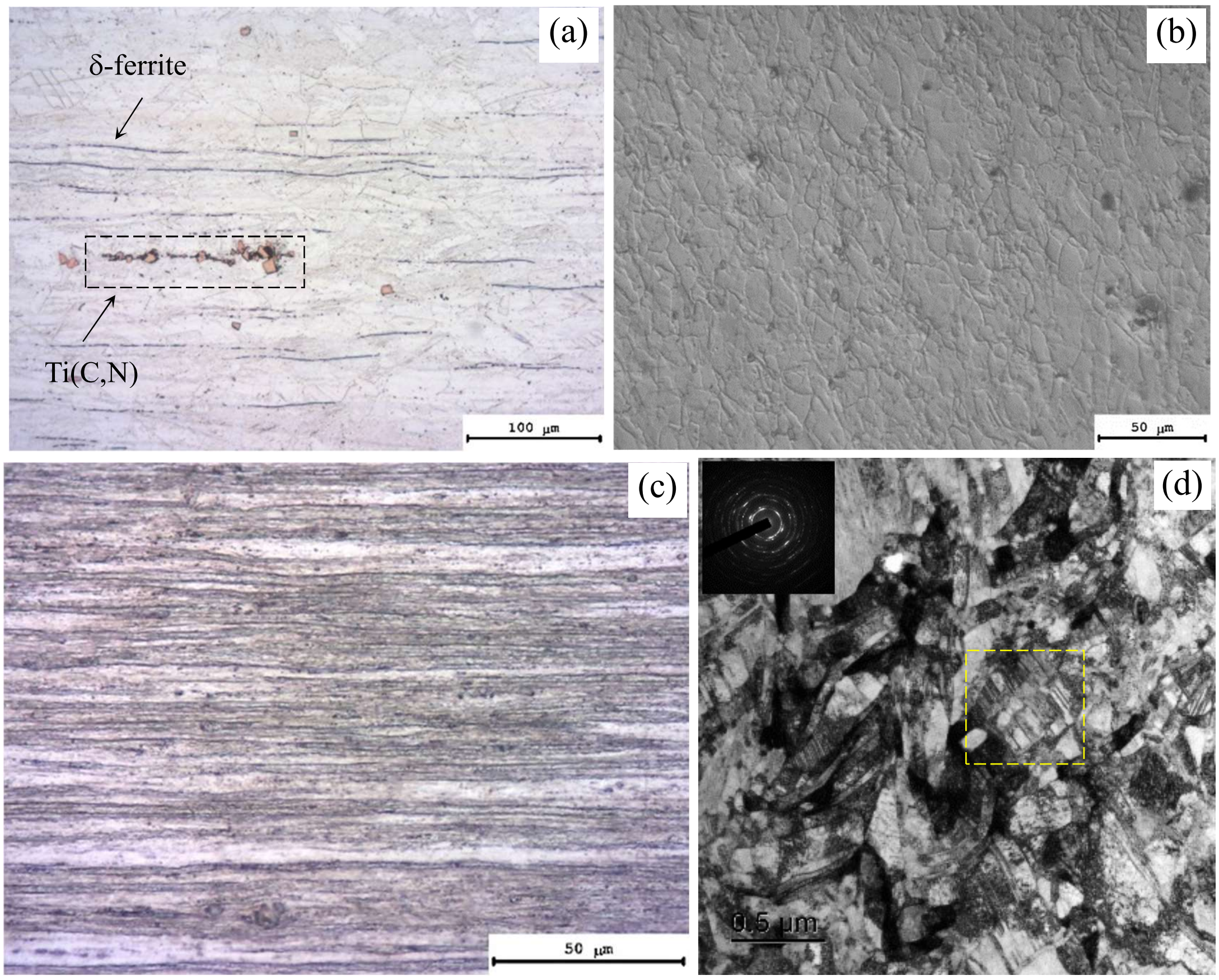

**Fig. 2** Microstructure of the central part of the rod in the initial state (a, b) and after RS (c, d). OM (a, b, c) and TEM (d)

Fig. 3 demonstrates the XRD patterns (using Cu radiation) of the coarse-grained steel. The diffraction patterns of the samples in the initial state contain maxima of the γ-Fe phase (PDF #01-071-4407, ICSD #53449) and the α-Fe phase (PDF #00-006-0696, ICSD #53802), which is obviously δ-ferrite. The intensities of the (111) γ-Fe and (222) γ-Fe maxima do not correspond to the expected ones (according to the PDF cards). This may indicate the presence of a texture (preferential orientation of the crystallites), which complicates the correct determination of the amount of δ-ferrite. The unit cell volumes of γ-Fe (46.51 Å$^3$) and α-Fe (23.94 Å$^3$) calculated by the Rietveld method exceed the tabulated values taken from the cif-files of the ICSD (2016): 23.54 Å$^3$ for α-Fe and 48.80 Å$^3$ for γ-Fe. The observed differences are probably related to the influence of an increased concentration of alloying elements affecting the unit cell parameters of the BCC and FCC phases of iron.

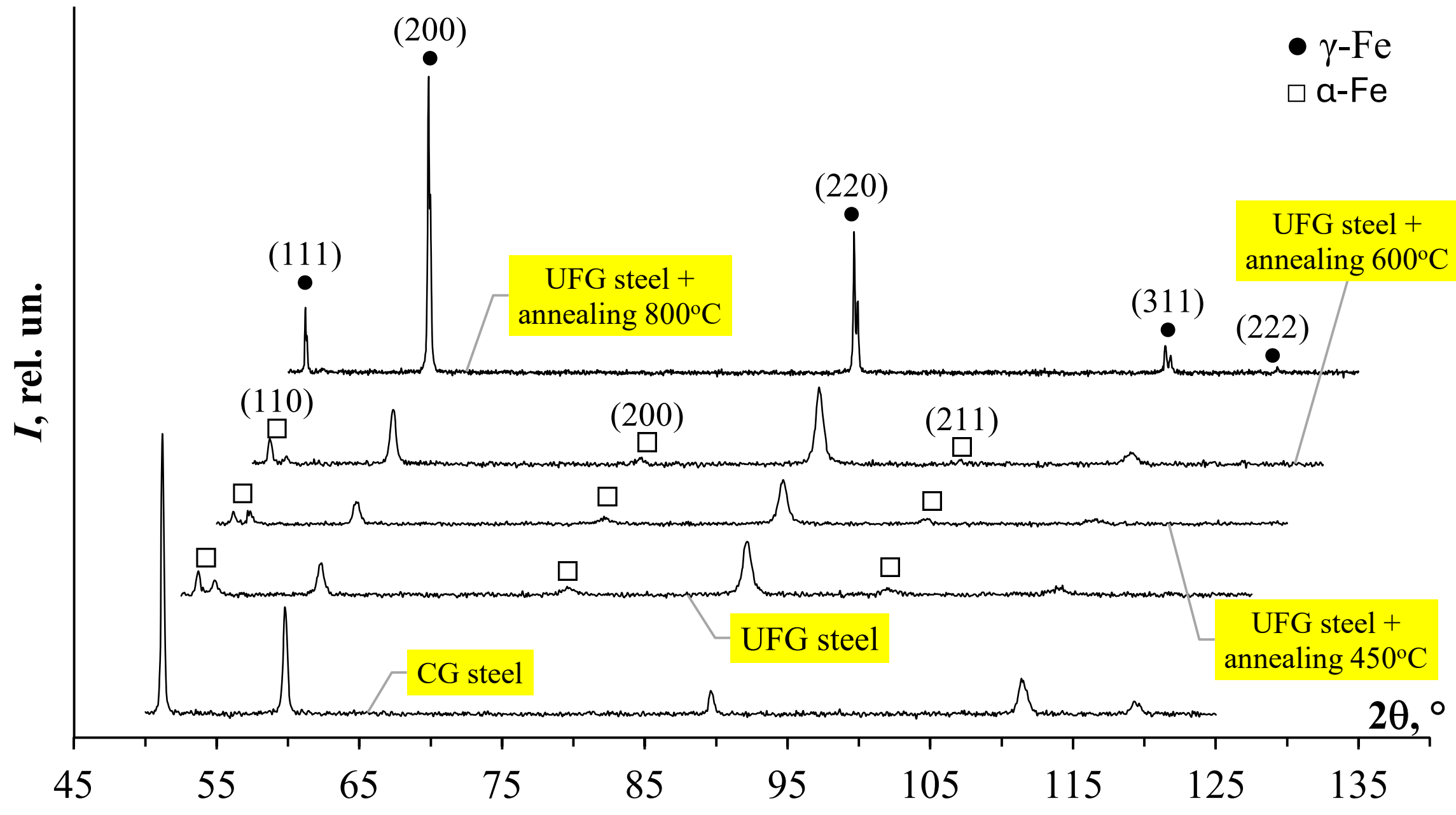

**Fig. 3** Results of XRD studies of the phase composition of steel 321

In the as-received state, coarse-grained steel has low microhardness (Hv = 1.8–1.9 GPa). The cross-section of the rod has a uniform hardness distribution (Fig. 9a). In the tensile tests, the steel in its initial state has yield tensile strength YTS = 510 MPa, ultimate tensile strength UTS = 680 MPa, and elongation to failure El = 45% (Table 2). The tensile stress–strain curve of the CG steel has an extended stage of uniform deformation, which is typical for a highly plastic material. The yield strength of the CG steel in the compression tests is YCS = 530 MPa (Table 2).

*UFG steel after Rotary Swaging*

Following RS, the steel develops a highly deformed, textured microstructure (Fig. 2c). In the center of the rod cross-section, the average size of the austenite fragments is ~0.5 μm (Fig. 2d).

Nanotwins are visible in some austenite grains, highlighted by the dotted line in Fig. 2d. During RS, partial refinement of δ-ferrite particles and fracture of Ti(C,N) particles occur, which are clearly visible on the polished surface of the section (Fig. 4). In some regions, δ-ferrite particles are displaced at an angle of 45°, probably due to the passage of a macro-shear band. In Fig. 3c, such a displacement of a δ-ferrite particle is indicated by the dotted line.

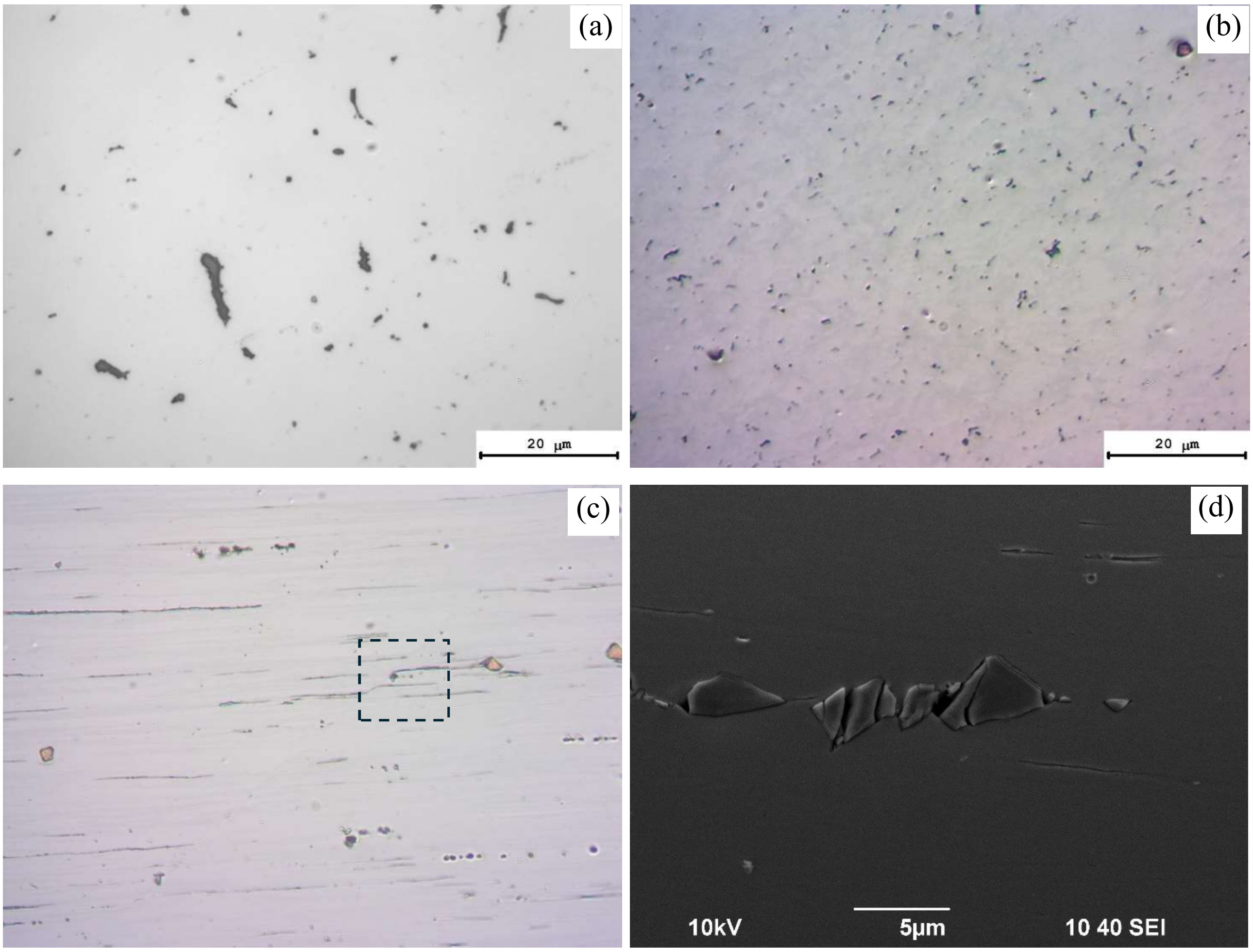

**Fig. 4** Particles of δ-ferrite (a, b, c) and Ti(C,N) (d) in the steel in the CG steel (a) and UFG steel after RS (b, c, d). Cross-section (a, b) and longitudinal sections (c, d) of the rod. OM (a, b, c) and SEM (d)

The results of the XRD indicate that maxima corresponding to γ-Fe (PDF #01-071-4407, ICSD #53449) and α-Fe (PDF #00-006-0696, ICSD #53802) are presented for the diffractogram of the UFG steel (Fig. 3). Cold RS leads to an increase in the intensity of the α-Fe diffraction peaks, indicating a higher volume fraction of strain-induced martensite in the UFG steel. The unit cell volume of α′-martensite in the deformed steel is 23.80–23.86 Å$^3$, which is 0.10–0.21% less than the volume of the BCC phase (δ-ferrite) in the undeformed coarse-grained steel.

The results of the MFA indicate that increasing the strain at RS leads to a increase content of the ferromagnetic phase in the steel rod (Fig. 5a). The average ferromagnetic phase content in the UFG rod ($D$ = 6 mm) is 12.3 ± 1.6%. The ferromagnetic phase content reaches 11% in the central part of the rod with a diameter of 6 mm. In the near-surface region, this proportion rises to 22%.

In the following, we will examine the effect of annealing temperature on the microstructure and phase composition of the UFG steel 321. This issue has been discussed in detail in [14–21], so we will focus only on new findings that are important for further analysis of the results of corrosion tests of UFG steel 321.

Fig. 5b illustrates the dependence of the martensite content $f_V(\alpha)$ on the temperature of 30-minute annealing of the UFG steel samples. This dependence exhibits a distinct two-stage behavior. At lower annealing temperatures (below 450°C), the martensite content remains unchanged. Increasing the annealing temperature from 500°C to 600°C leads to a reduction in the fraction of martensite from 11.3% to 6%. The XRD pattern of the sample annealed at 600°C reveals only low-intensity α-Fe (200) and (211) diffraction peaks. Annealing at 650°C completes the reverse transformation of martensite into austenite, with no detectable martensite remaining in the specimens, as confirmed by both magnetic phase analysis and XRD.

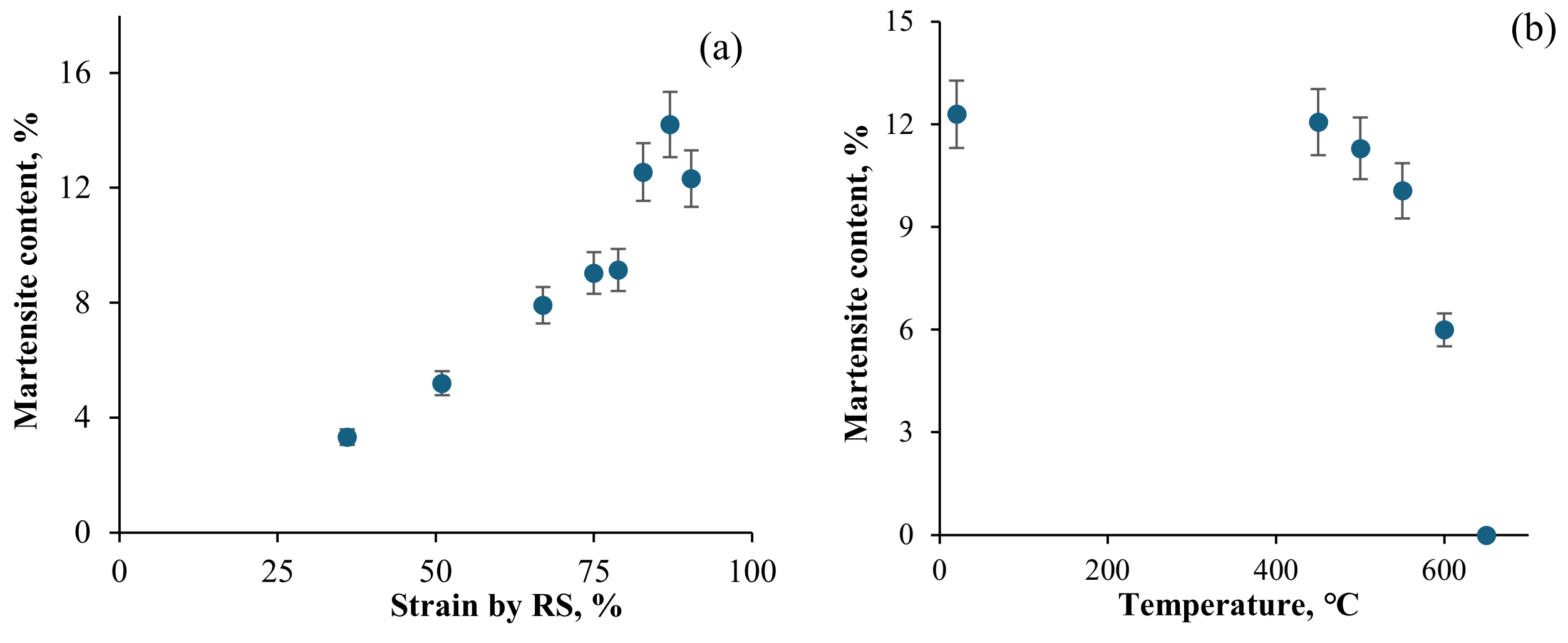

**Fig. 5** Dependence of the martensite consist $f_V(\alpha)$ in UFG steel on the strain by RS (a) and on the annealing temperature (b)

The optic and SEM studies of the microstructure reveal that the formation of nano- and submicron secondary particles occurs in the temperature range of 450–500°C (Fig. 6). Recrystallization processes are not observed. The results of the EDS microanalysis indicate that the precipitated particles contain Ti (Fig. 6a) and C (Fig. 6b). This finding suggests that the precipitated secondary particles in the annealed sample are titanium carbide (TiC). This conclusion is consistent well with the data presented in [16], which also reported the precipitation of TiC particles during annealing of the UFG steel. The microstructure of the steel remained stable after annealing at 450–500°C, and no start of recrystallization were observed using by optical metallography or SEM.

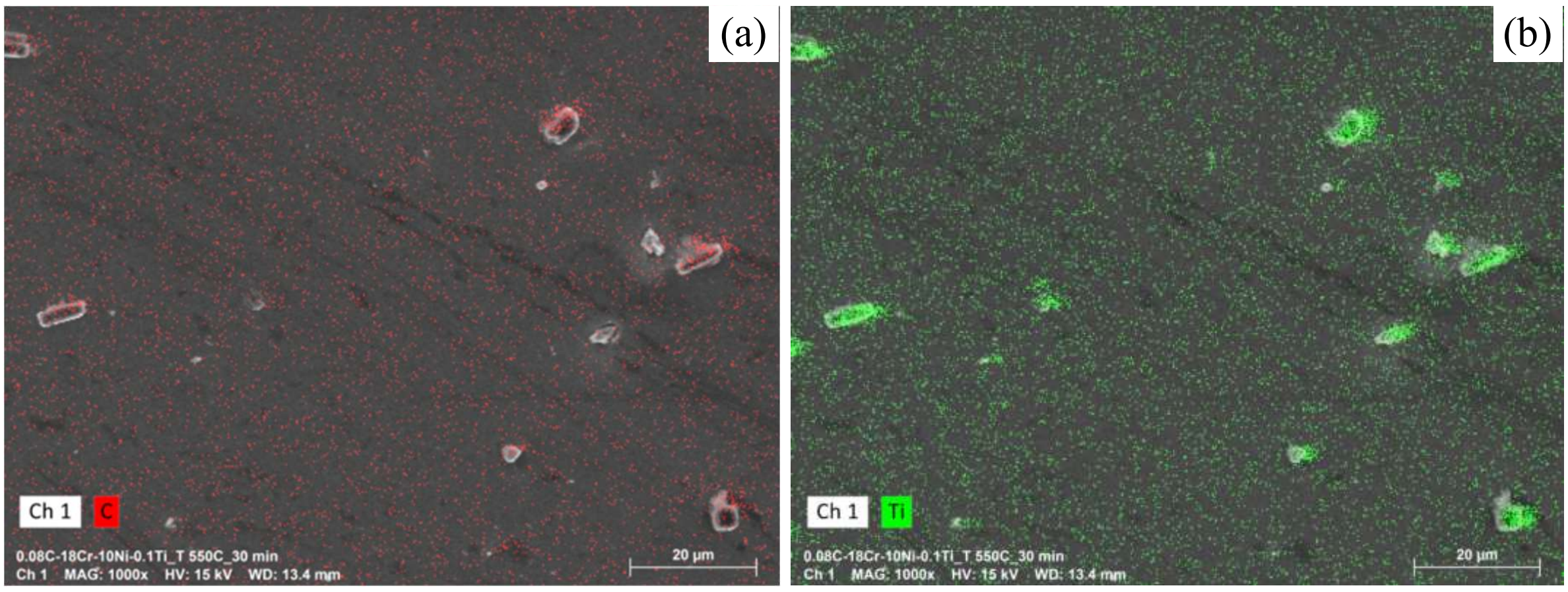

**Fig. 6** Results of the EDS microanalysis of the composition of TiC secondary particles precipitated during annealing at 550°C, 30 min. SEM [40]

Annealing at 650°C initiates the recrystallization process, resulting in the nucleation of the first recrystallized austenite grains within the highly deformed microstructure. The regions of recrystallized microstructure in Fig. 7a are highlighted with a yellow dashed line. After annealing at 650°C, the volume fraction of the recrystallized phase reaches 10%, with an average austenite grain size of 3–5 μm (Fig. 7a). Increasing the annealing temperature to 750°C results in a higher volume fraction of recrystallized grains., but the average size of the recrystallized grains remains relatively unchanged (Fig. 7b). After annealing at 800°C, the rod has a fully recrystallized microstructure, and the average size of the recrystallized austenite grains in the surface of the rod is ~1.5 times larger than that of the average grains in the central region of the rod (Figs. 7c–d).

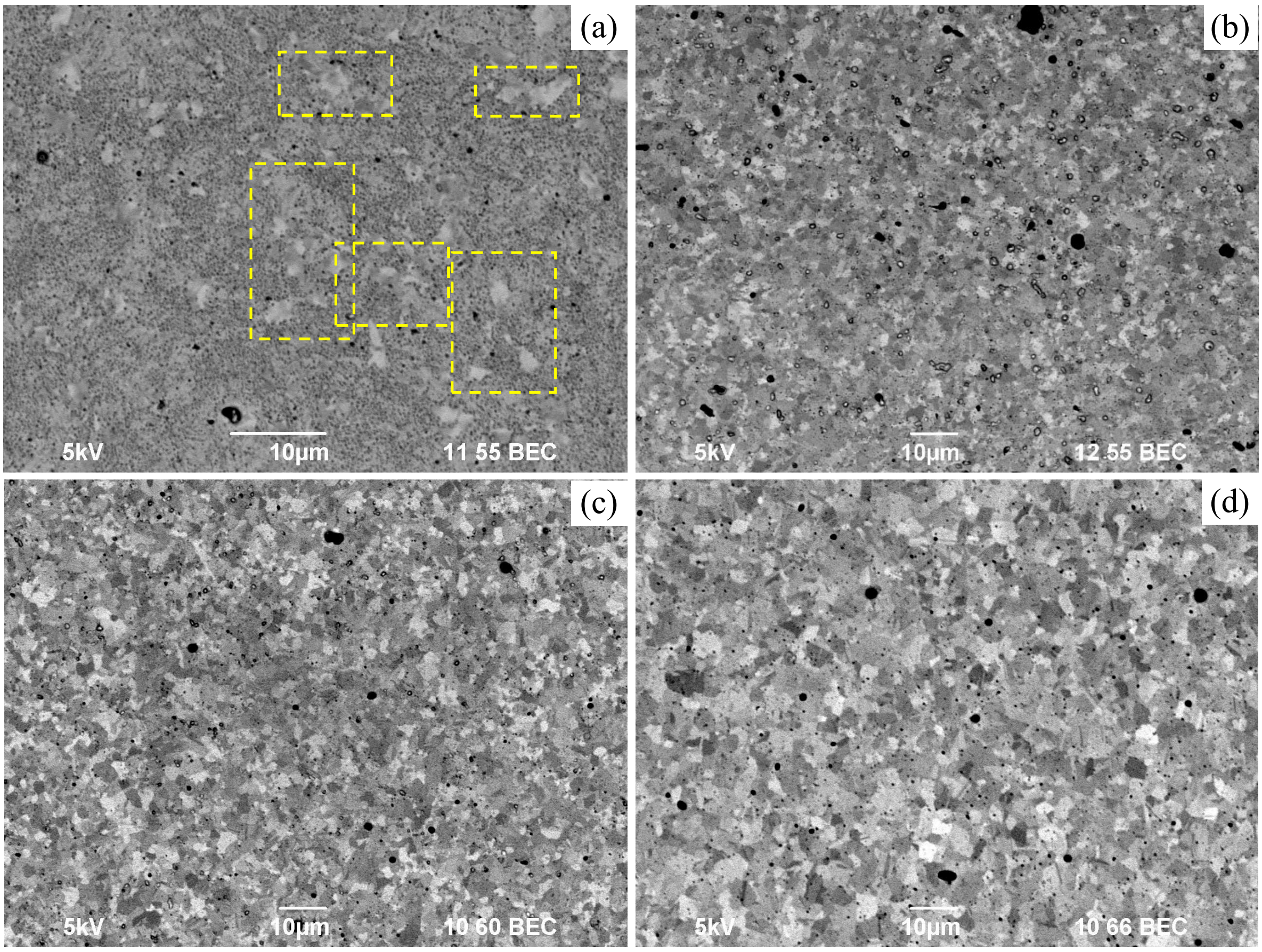

**Fig. 7** SEM images of the austenite microstructure of the central part (a, b, c) and surface (d) of the UFG rod after annealing at 650°C (a), 750°C (b), and 800°C (c, d). Cross-section of the rod [40]

After annealing at 800°C, the diffraction patterns of the recrystallized steel contain high-intensity γ-Fe peaks, while the intensity of the α-Fe peaks is very low and comparable to the background intensity. Highly fragmented δ-ferrite particles are visible on the polished metallographic sections. This leads to the conclusion that the low-intensity α-Fe peaks correspond to δ-ferrite.

Below is a brief description of the mechanical properties of UFG steel 321.

RS results in a significant increase in the YTS and UTS of steel. After RS, the UTS of steel reaches 1580 ± 30 MPa, while the El decreases to 4.4 ± 0.4% (Table 2). The σ(ε) curves do not exhibit a stage of uniform plastic flow (the uniform strain remains below 0.1–0.3%). Analysis of the compression test results shows that the formation of the UFG microstructure by RS leads to an increase in the yield strength (YCS) from 530 ± 15 to 1040 ± 25 MPa (Table 2).

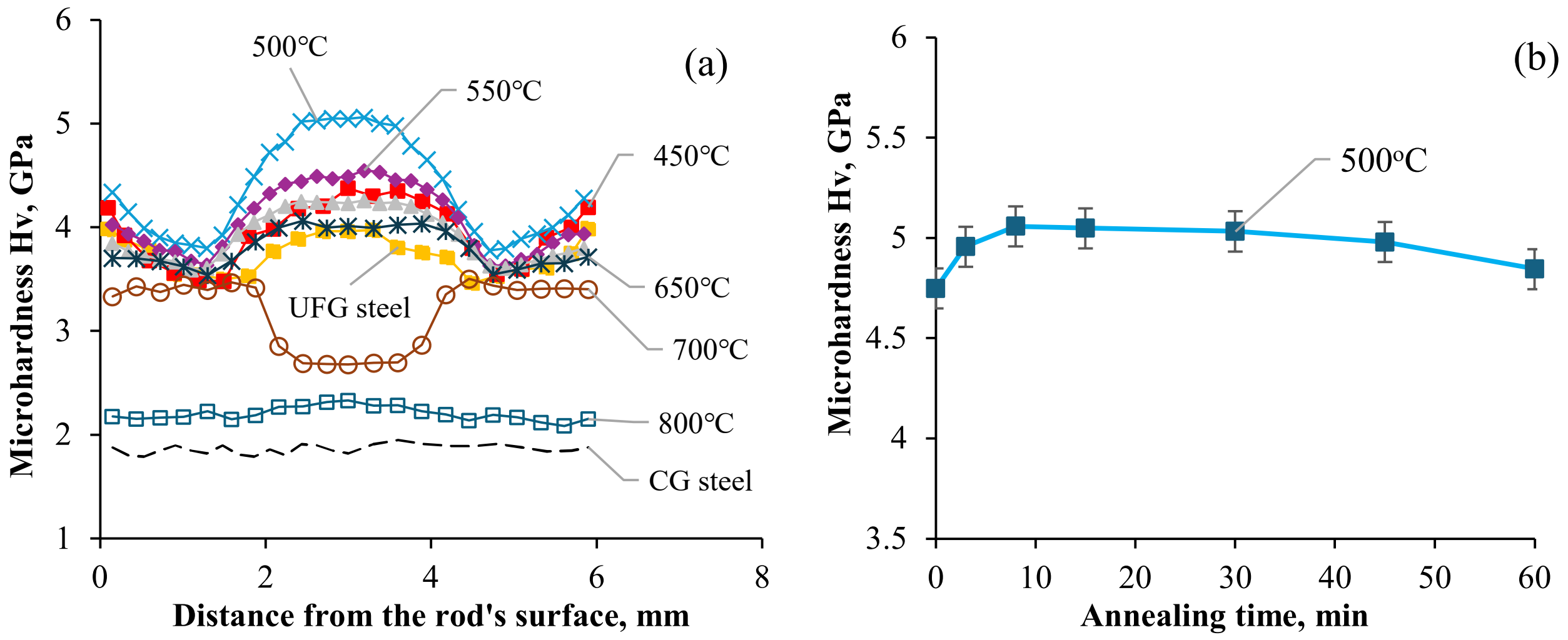

**Fig. 8** Results of mechanical tests of steel 321 in different states: (a) microhardness distribution graphs in the cross-section; (b) the dependence of the microhardness at the rod center on the annealing time at 500°C. The annealing temperatures of the UFG steel are indicated next to the graphs

Fig. 8a illustrates the microhardness distribution graph in the cross-section of the steel 321 rod in different states. In the initial coarse-grained state, the rod exhibits a uniform Hv distribution. However, after RS, the nature of the microhardness distribution changes fundamentally. The microhardness profile across the cross-section of the steel rod exhibits two peaks and a single minimum. In the center of the rod and on its surface, the Hv value reaches 3.96 ± 0.09 and 3.99 ± 0.08 GPa, while at a distance of 1–2 mm from the surface, the Hv decreases to ~ 3.40 ± 0.08 GPa.

**Table 2.** Mechanical properties of steel 321

| State | Tensile tests | | | | Compression tests | Hardness Hv, GPa | |
|---|---|---|---|---|---|---|---|
| | YTS, MPa | UTS, MPa | El, % | ψ, % | YCS, MPa | Center | Surface |
| CG | 510 ± 20 | 680 ± 20 | 45 ± 5 | – | 530 ± 15 | 1.8–1.9 | |
| UFG | 1550 ± 30 | 1580 ± 30 | 4.4 ± 0.4 | 34 ± 4 | 1040 ± 25 | 3.96 ± 0.09 | 3.99 ± 0.08 |
| RS + 450°C | 1695 ± 35 | 1770 ± 40 | 2.8 ± 0.3 | 32 ± 4 | 1225 ± 25 | 4.38 ± 0.08 | 4.19 ± 0.09 |
| RS + 500°C | 1535 ± 30 | 1545 ± 35 | 2.7 ± 0.3 | 46 ± 4 | 1330 ± 30 | 5.04 ± 0.08 | 4.28 ± 0.10 |
| RS + 550°C | 1535 ± 30 | 1540 ± 30 | 3.8 ± 0.4 | 32 ± 4 | 1240 ± 30 | 4.03 ± 0.08 | 3.94 ± 0.08 |
| RS + 600°C | 1390 ± 30 | 1415 ± 35 | 4.8 ± 0.5 | 39 ± 4 | 1040 ± 25 | 3.94 ± 0.12 | 3.80 ± 0.07 |
| RS + 650°C | 1315 ± 30 | 1315 ± 30 | 5.0 ± 0.5 | 51 ± 5 | 1040 ± 25 | 3.70 ± 0.11 | 3.72 ± 0.08 |
| RS + 700°C | 1000 ± 25 | 1080 ± 25 | 10 ± 2 | 61 ± 7 | 950 ± 25 | 2.69 ± 0.11 | 3.40 ± 0.06 |
| RS + 750°C | 510 ± 20 | 770 ± 20 | 24 ± 3 | 71 ± 7 | 555 ± 20 | 2.34 ± 0.10 | 2.34 ± 0.06 |
| RS + 800°C | 430 ± 20 | 755 ± 20 | 32 ± 3 | 72 ± 8 | 420 ± 20 | 2.28 ± 0.07 | 2.15 ± 0.05 |

Annealing at 450–500°C results in an increase in the microhardness: Hv at the center of the rod's cross-section reaches 4.38 ± 0.08 GPa, while Hv is 4.19 ± 0.09 GPa at the surface of the rod; the microhardness at a distance of 1–2 mm from the rod surface remains virtually unchanged and is 0.43–0.47 GPa. A slight increase in Hv is probably associated with the processes of polygonization and the formation of dislocation boundaries. The ultimate tensile strength after annealing at 450°C increases to UTS = 1770 ± 40 MPa, while the El decreases to 2.8 ± 0.3%. The yield strength in tensile and compressive tests after annealing at 450°C increases to 1695 ± 35 MPa (ΔYTS ~ +145 MPa) and 1225 ± 25 MPa (ΔYCS ~ +185 MPa), respectively.

The ultimate tensile strength after annealing at 450°C increases to UTS = 1770 ± 40 MPa, while the El decreases to 2.8 ± 0.3%. The yield strength in tensile and compressive tests after annealing at 450°C increases to 1695 ± 35 MPa (ΔYTS ~ +145 MPa) and 1225 ± 25 MPa (ΔYCS ~ +185 MPa), respectively.

After annealing at 500°C, the microhardness of the rod center reaches its maximum value (~5.04 ± 0.08 GPa). The microhardness of the surface increases to 4.28 ± 0.10 GPa, and the yield strength in compression tests increases to 1330 ± 30 MPa (Table 2, Fig. 8a). Notably, the mechanical properties determined in tensile tests begin to decrease after annealing at this temperature.

A further increase in the temperature leads to a decrease in the microhardness (Fig. 8a) and the mechanical properties determined in tensile and compressive tests (Table 2). It is interesting to note that the microhardness of the rod center decreases much more rapidly when heated than the microhardness of the surface layer. After annealing at 700°C, the microhardness distribution across the cross-section reverses, resulting in a lower microhardness at the rod center compared to its surface. This finding suggests that recrystallization proceeds more rapidly in the sample center compared to the surface layer. After annealing at 750–800°C, a uniform microhardness distribution is observed across the sample cross-section. The microhardness of the rod annealed at 800°C is 2.15–2.28 GPa and slightly exceeds the microhardness of the coarse-grained steel (1.8–1.9 GPa). Further increasing the annealing temperature to 800°C leads to an increase in the El of the steel and a decrease in the UTS and YTS. After annealing at 800°C, the mechanical properties of the UFG steel become comparable to those of the coarse-grained steel in the as-received state (Table 2).

Fig. 8b demonstrates that the microhardness at the sample center, as a function of isothermal annealing time at 500°C, exhibits a non-monotonic trend with a maximum. This indirectly supports the hypothesis of secondary TiC nanoparticle precipitation during annealing (Fig. 6).

It is important to note that the effect of increasing the tensile strength and microhardness during low-temperature annealing of UFG steel has previously been observed by other researchers. A similar result was previously reported in [16], where the metastable austenitic steel was processed by multi-stage RS with a temperature reduction from 450 to 20°C. This enabled the authors of [16] to produce UFG steel with a high content of α'-martensite (68% vol.) and a favorable combination of mechanical properties (UTS = 1304 MPa at El = 10.5%). Annealing at 475°C resulted in an increase in the martensite content to 90% and the precipitation of TiC nanoparticles. As a result, the ultimate strength of the UFG steel increased to UTS = 1939 MPa, while the elongation to failure decreased to El = 8%. The authors of [16] did not disclose the probable reasons for the increase in martensite content from 68 to 90% during low-temperature annealing. Therefore, we attribute this additional increase in UTS primarily to the precipitation of TiC nanoparticles.

## 3.2 Corrosion tests

### *3.2.1 General tests*

Fig. 9a presents the $E(t)$ curves for the UFG steel samples after RS and after annealing at 800°C. The tests were conducted in open-circuit mode for 1 h. The average error in determining the open-circuit potential (± 6 mV) was calculated based on the analysis of triplicate studies of three identical samples.

From Fig. 9a it can be seen that at the initial moment there is a rapid decrease in the open-circuit potential. The minimum potential values of the unannealed samples cut from the center of the rod are achieved after exposure for ~5–6 min ($E_{min}$ = –400 ± 6 mV (vs. Ag/AgCl)), and for samples cut from the surface layer of the rod – after 10–12 min of exposure ($E_{min}$ = –390 ± 6 mV (vs. Ag/AgCl)). After 1 h of exposure, the open-circuit potential value for the unannealed UFG samples reaches its steady-state value of –372 ± 6 mV (vs. Ag/AgCl). There is no difference between the potentials of the central part of the rod ($E_C$) and the surface of the rod ($E_S$). Recrystallization annealing results in a slight decrease in $E_C$ by ~20 mV and an increase in $E_S$ by ~10–15 mV. After 1 h of exposure in a corrosive environment, the potential of the surface layer in the annealed samples has not yet reached its steady-state value and continues to increase slowly.

Fig. 9b presents the open-circuit potential values as a function of the annealing temperature. Fig. 9b shows that an increase in the annealing temperature leads to an increase in the difference between the potentials of the center and the surface of the UFG rod. After annealing at 700–800°C, accompanied by the formation of an equiaxed fine-grained austenitic structure (Fig. 7), the difference reaches $\Delta E = E_S - E_C$ = 35–50 mV (Fig. 9b). Thus, in recrystallized steels, there is a small potential difference between the centers and the surface of the rod, with the surface of the rod having higher open-circuit potential values.

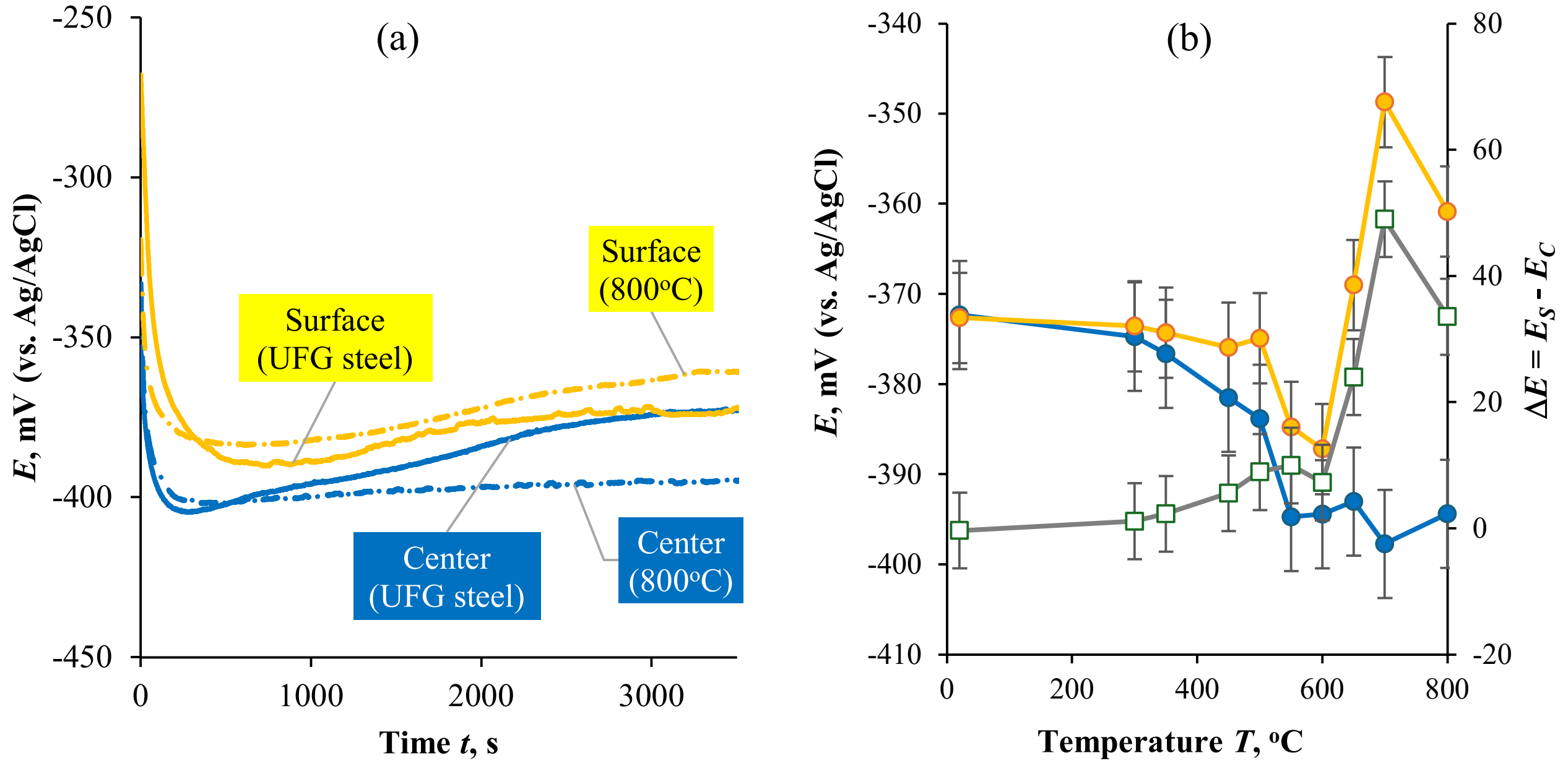

**Fig. 9** Test results in open-circuit mode: (a) potential-time curves $E(t)$ for unannealed UFG samples (solid lines) and UFG samples after annealing at 800°C (dotted lines); (b) the dependence of the open-circuit potential on the annealing temperature for samples from the surface ($E_S$) and the center of the rod ($E_C$), as well as the difference $\Delta E = E_S - E_C$. In Fig. 9a, samples cut from the center of the rod are marked with solid lines; samples cut from the surface layer are marked with dotted lines

Fig. 10a presents a typical view of potentiodynamic polarization curves $j(E)$ of the current density–potential for the UFG steel samples in the post-RS condition and after annealing at 800°C. Fig. 10 shows that the curves $j(E)$ have a cathodic region (region #1), an active corrosion region (region #2), a passive potential region (region #3), and a transpassive region (region #4). The active corrosion region was characterized by the value of the critical current density $j_{acr}$ (Fig. 10b). The passivation region was characterized by the passivation potential ($E_{pass}$), the potential of the end of the passivation region and the beginning of the pitting region ($E_{pi}$), the width of the passivation region ($\Delta E_P = E_{pp} - E_{pass}$) and the critical current density in the passive region $j_{pcr}$. The transpassive region begins at potentials of ~1000–1070 mV (vs. Ag/AgCl), and this region has not been analyzed by us, since the steel under study is not intended for operation in this potential range. The parameters of the potentiodynamic polarization curves $j(E)$ are presented in Table 3.

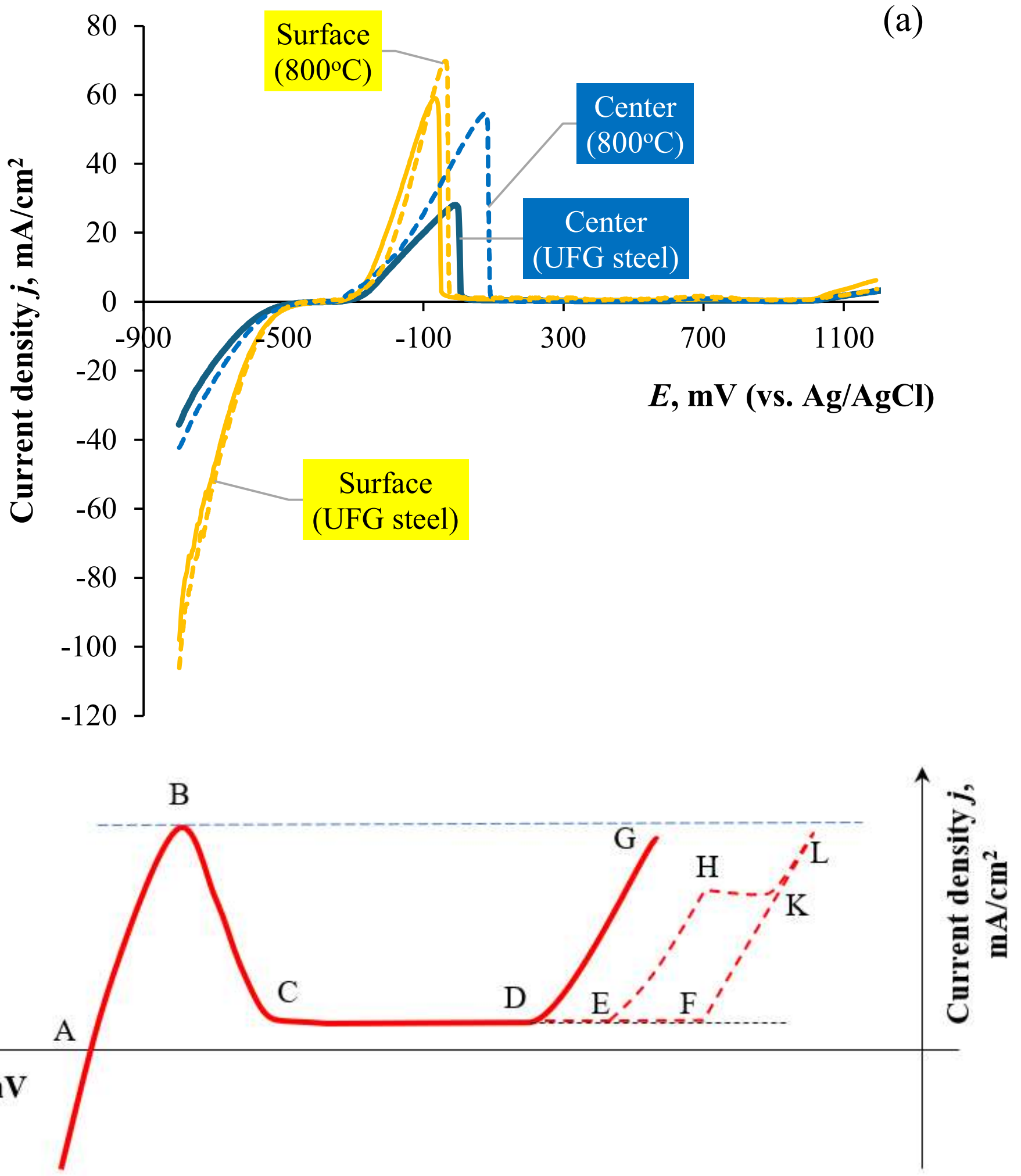

**Fig. 10** Potentiodynamic polarization curves $i(E)$ for steel 321: (a) curves for unannealed UFG samples (solid lines) and UFG samples after annealing at 800°C (dashed lines); (b) a qualitative view of the $i$(E) dependence. Samples cut from the center of the rod are marked with solid lines; samples cut from the surface layer – with dashed lines

**Table 3.** Results of the analysis of potentiodynamic polarization curves for steel 321

| State | Surface | | | | | | | | | Center | | | | | | | | |
|---|---|---|---|---|---|---|---|---|---|---|---|---|---|---|---|---|---|---|
| | Region #1 (Tafel test) [1] | | | Region #2 | | Region #3 | | | | Region #1 (Tafel test) [1] | | | Region #2 | | Region #3 | | | |
| | | | | (General test) [2] | | | | | | | | | (General test) [2] | | | | | |
| | $E_{corr}$, mV | $j_{corr}$, mA/cm$^2$ | $V_{corr}$, mm/year | $j_{acr}$, mA/cm$^2$ | $E_{acr}$, mV | $E_{pass}$, mV | $E_{pp}$, mV | $\Delta E_P$, mV | $j_{pcr}$, mA/cm$^2$ | $E_{corr}$, mV | $j_{corr}$, mA/cm$^2$ | $V_{corr}$, mm/year | $j_{acr}$, mA/cm$^2$ | $E_{acr}$, mV | $E_{pass}$, mV | $E_{pp}$, mV | $\Delta E_P$, mV | $j_{pcr}$, mA/cm$^2$ |
| After RS | -380 | 0.10 | 0.77 | 57 | -59 | -49 | 1002 | 1051 | 0.5 | -368 | 0.12 | 0.92 | 28 | -22 | 3 | 1050 | 1047 | 0.3 |
| RS + 450°C | -382 | 0.11 | 0.84 | 48 | -70 | -43 | 1030 | 1046 | 1.6 | -381 | 0.06 | 0.44 | 23 | -34 | -19 | 1015 | 1034 | 0.3 |
| RS + 500°C | -377 | 0.09 | 0.66 | 45 | -89 | -58 | 1034 | 1092 | 1.1 | -384 | 0.06 | 0.43 | 22 | -37 | -21 | 1014 | 1035 | 0.3 |
| RS + 550°C | -376 | 0.09 | 0.72 | 53 | -34 | -18 | 1027 | 1045 | 0.9 | -394 | 0.11 | 0.81 | 61 | -7 | 21 | 1006 | 1027 | 0.9 |
| RS + 600°C | -386 | 0.09 | 0.66 | 47 | -43 | -9 | 1027 | 1036 | 0.7 | -392 | 0.06 | 0.48 | 54 | 27 | 22 | 1006 | 1028 | 0.6 |
| RS + 650°C | -381 | 0.09 | 0.59 | 32 | -73 | -45 | 1018 | 1063 | 1.0 | -402 | 0.07 | 0.55 | 62 | 24 | 64 | 1017 | 953 | 0.8 |
| RS + 700°C | -378 | 0.08 | 0.64 | 36 | -86 | -64 | 1030 | 1094 | 1.1 | -392 | 0.05 | 0.41 | 56 | 95 | 132 | 1011 | 879 | 0.8 |
| RS + 800°C | -369 | 0.07 | 0.56 | 69 | -38 | -23 | 1014 | 1037 | 0.9 | -391 | 0.05 | 0.42 | 54 | 80 | 98 | 1008 | 910 | 0.7 |

Notes:

[1] The average error in determining the values of $E_{corr}$, $j_{corr}$ and $V_{corr}$ is ± 5 mV (vs. Ag/AgCl), ± 0.01 mA/cm$^2$ and 0.02 mm/year, respectively.

[2] The average error in determining the values of $j_{acr}$ and $E_{acr}$ is ± 2 mA/cm$^2$ and ± 2 mV (vs. Ag/AgCl), respectively. The average error in determining the values of $E_{pass}$, $E_{pp}$, $\Delta E_P$ and $j_{pcr}$ is ± 2 mV (vs. Ag/AgCl), ± 5 mV (vs. Ag/AgCl), ± 7 mV (vs. Ag/AgCl), ± 0.1 mA/cm$^2$, respectively.

It can be seen from Table 3 and Fig. 10 that for the samples cut from the surface of the rod, the maximum current density in the active corrosion region ($j_{acr}$) is almost twice as high as the maximum current density for the samples cut from the center of the rod. The position of the maximum in region #2 on the potentiodynamic polarization curves $i(E)$ for the samples cut from the center of the rod is shifted by 20–25 mV toward more positive potentials than the position of the maximum for the samples cut from the surface layer.

Table 3 shows that annealing leads to a smooth decrease in the $j_{acr}$ for the samples cut from the surface layer of the rod and an increase in the value of $j_{acr}$ for the samples cut from the center of the rod. The value of the potential $E_{acr} = E\ (j = j_{acr})$ for the metal surface of the rod remains practically unchanged, but $E_{acr}$ for the samples from the central part of the rod shifts from –20 mV to 80–95 mV.

The width of the passive region $\Delta E_P = E_{pp} - E_{pass}$ for the rod surface is practically independent of the annealing temperature. An increase in the annealing temperature leads to a decrease in the width of the passive region for the samples cut from the center of the rod by 140–150 mV and an increase in the passivation potential $E_{pass}$. The critical current density in the passive region $j_{pcr}$ is practically independent of the annealing temperature and differs slightly for the center and surface of the steel rod (Table 3).

### *3.2.2 Tafel test*

Fig. 11 shows the Tafel curves for the samples after RS and after annealing at 800°C. The curves have a typical appearance for electrochemical corrosion of austenitic steels. The parameters of the $i(E)$ curves (corrosion current density $j_{corr}$, corrosion potential $E_{corr}$) are presented in Table 3.

Table 3 shows that the corrosion potential $E_{\mathrm{corr}}$ of the rod surface is slightly higher than the $E_{\mathrm{corr}}$ value for the center of the rod. However, this difference only slightly exceeds the variability of properties (± 10 mV) observed when analyzing identical samples cut from different sections of the rod (the beginning, the center, and the end of the rod). The corrosion current density $j_{\mathrm{corr}}$ on the surface and at the center of the rod differs by 0.02 mA/cm$^2$, which is comparable to the error in determining this value (± 0.01 mA/cm$^2$). Thus, it can be concluded that the Tafel method is insensitive to differences in the characteristics between the center and surface of the rods after RS.

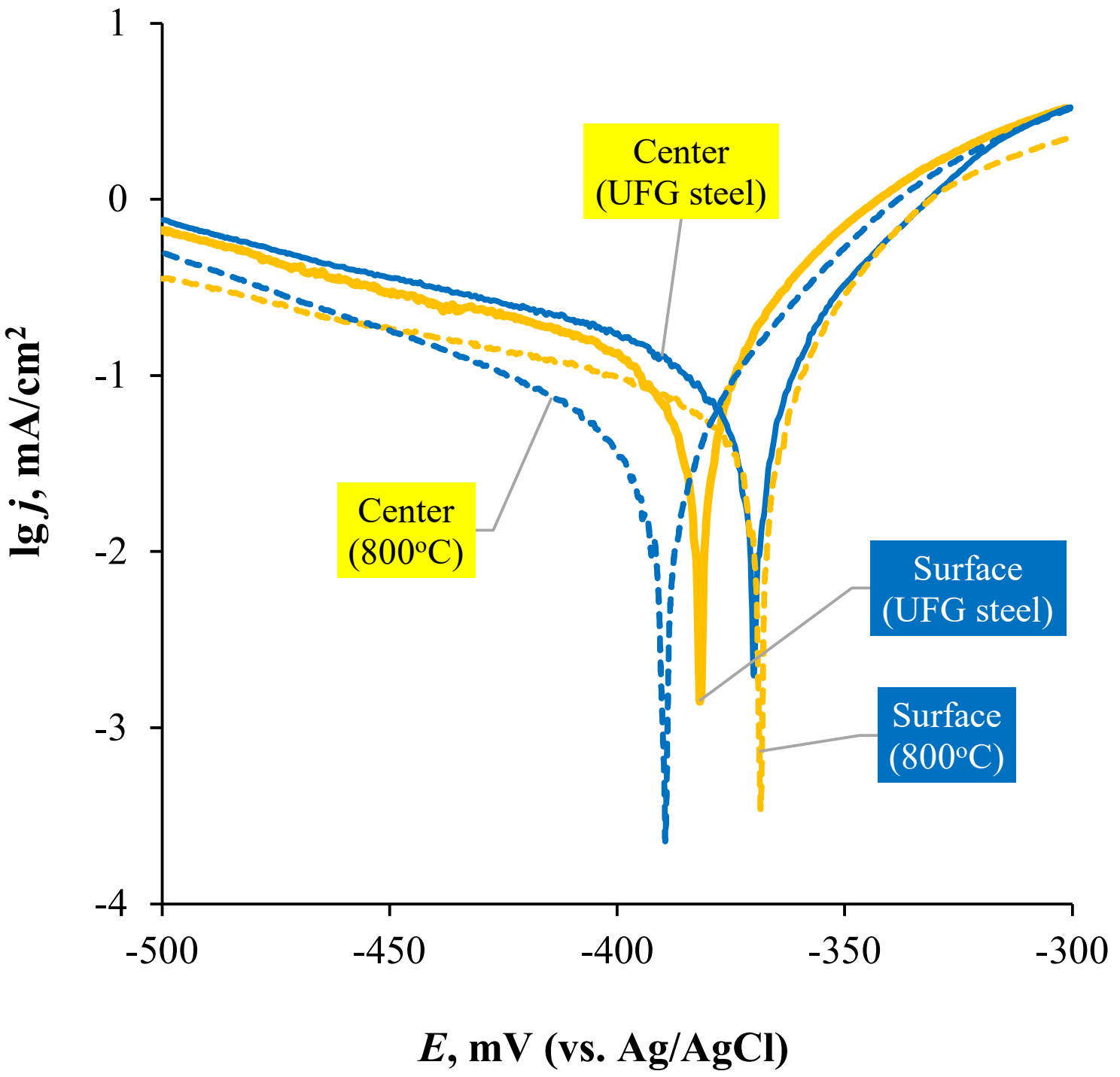

**Fig. 11** Curves of $i(E)$ for the transition region (from region #1 to region #2). Fig. 11 shows the results of studies of the unannealed UFG samples (solid lines) and UFG samples after annealing at 800°C (dashed lines)

Annealing does not lead to significant changes in the corrosion potential $E_{corr}$ of the rod surface, and the corrosion current density of the rod surface decreases slightly by 0.03–0.04 mA/cm$^2$. This corresponds to a reduction in the corrosion rate by 0.21–0.26 mm/year. The corrosion potential of the central part of the rod increases slightly by 23–32 mV, while the corrosion current density of the central part of the rod decreases by 0.07 mA/cm$^2$ (Table 3). This reduction in the corrosion current density corresponds to a decrease in the corrosion rate from 0.92 to 0.41–0.42 mm/year. On the surface of the samples after electrochemical studies, traces of mechanical grinding and polishing are visible, which indirectly indicate the absence of intensive corrosion (Fig. 12). The most intense corrosion damage on the sample surfaces is observed near non-metallic inclusions and δ-ferrite particles.

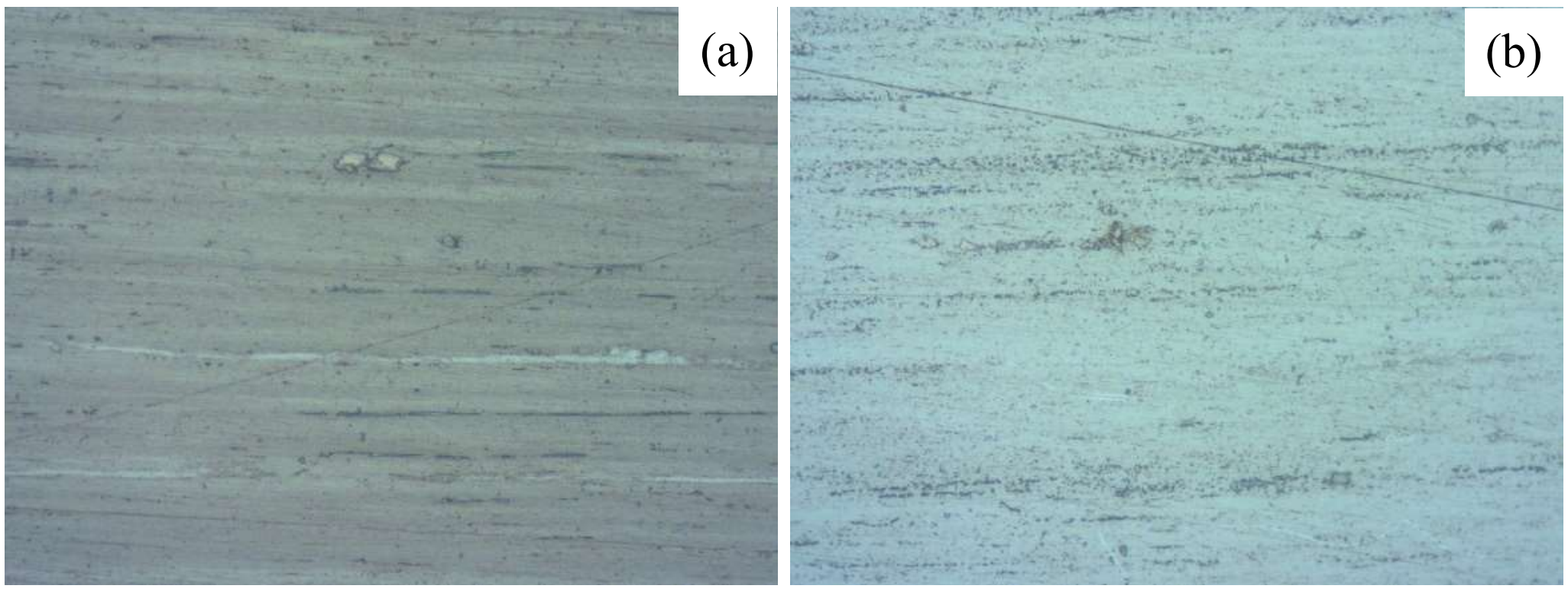

**Fig. 12** The surface of UFG steel samples after electrochemical tests using the Tafel test method: (a) unannealed steel; (b) steel annealed at 700°C. Optical microscopy

*3.2.3 Pitting resistance test*

Fig. 13a shows the typical current – stationary exposure time curves $i(t)$ for the samples of UFG steel at potential values of 600 mV and 1200 mV (vs. Ag/AgCl). Fig. 13b presents a comparative analysis of the effect of annealing temperature on the $i(t)$ curves at the potential $E$ = 1200 mV (vs. Ag/AgCl). As already noted in the Materials and Methods section, a potential value of 600 mV corresponds to the passivation region, and a potential of 1200 mV corresponds to the beginning of the repassivation region (Fig. 10). Before applying potentials of 600 and 1200 mV, the samples were briefly exposed to a potential of –500 mV for 2 min. The curve $i(t)$ in Fig. 13a is shown by the dashed line. For ease of analysis, the $i(t)$ curves are plotted in relative coordinates $i/i_0(t)$, where $i_0$ is the current at the initial moment ($t$ = 0 s).

Fig. 13a shows that increasing the exposure time at potentials of 600 and 1200 mV leads to a decrease in the current. It should be noted that in the passivation region, at $E$ = 600 mV, the current reaches its steady-state value within 10–20 s, while at $E$ = 1200 mV, it takes ~4–5 min to reach the steady state. Fig. 13b shows that this type of $i(t)$ dependence is typical for all samples, regardless of their annealing temperature.

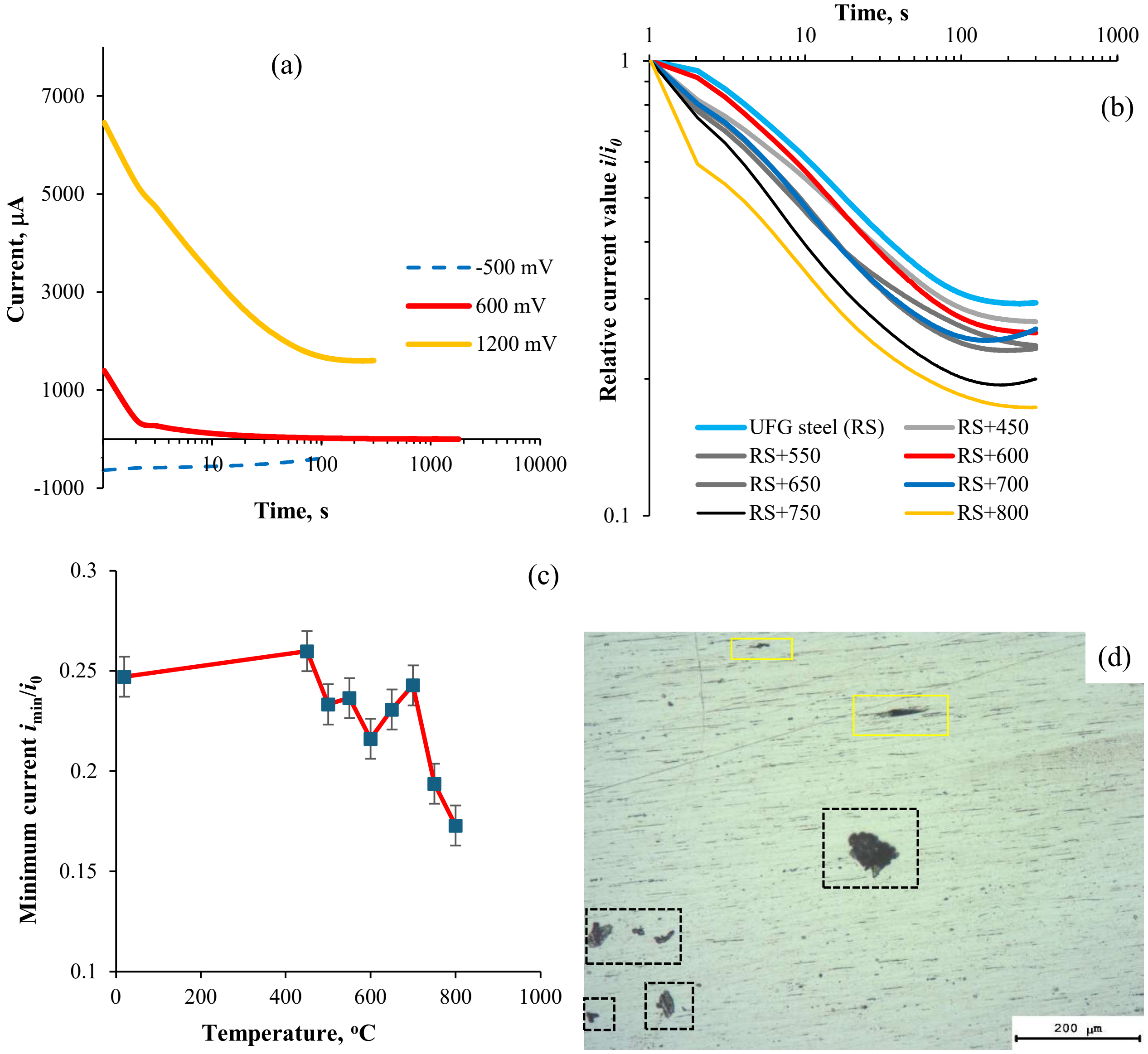

**Fig. 13** Test results of UFG steel to PC susceptibility: (a) $i(t)$ curves at various potential values for unannealed UFG steel; (b) $i/i_0$–$t$ curves for samples annealed at different temperatures, presented in logarithmic form; (c) the dependence of the relative passivation current $i_{min}/i_0$ on the annealing temperature; (d) an optical microscopy image of the surface of an annealed UFG steel sample after exposure for 60 min at a potential of 1200 mV (vs. Ag/AgCl) (comments in the text).

It is interesting to note that an increase in the annealing temperature leads to a non-monotonic change in the current $i_{min}$, which is achieved during the steady-state exposure at $E$ = 1200 mV. Fig. 13c shows that, starting from an annealing temperature of 600°C, the value of $i_{min}/i_0$ increases, and, after annealing at 800°C, the value of $i_{min}/i_0$ decreases by 58–60 % relative to the initial value. This result indirectly indicates that the fully annealed steel has increased resistance to pitting. Fig. 13g shows that three types of corrosion defects are observed on the surface of the annealed samples after steady-state exposure at $E = 1200$ mV: (i) large, equiaxed pitting defects on non-metallic inclusions; (ii) elongated defects on δ-ferrite particles; (iii) uniformly distributed small pitting defects, the formation of which is likely associated with the precipitation of Ti(C,N) particles. Large Type 1 pitting defects in Fig. 13d are marked with a black dashed line, and typical elongated Type 2 defects are marked with a yellow solid line (Fig. 13g). Note that the temperature at which an increase in $i_{min}/i_0$ is observed (Fig. 13c) is close to the temperature at which Ti(C,N) particles begin to precipitate (Figs. 6, 8b). It can be assumed that the non-monotonic change in the $i_{min}/i_0$ value during annealing of the UFG steel is due to two simultaneous processes: the precipitation of Ti(C,N) nanoparticles increases the UFG steel's susceptibility to pitting corrosion, while a reduction in the size of δ-ferrite particles decreases the steel's susceptibility to pitting corrosion.

*3.2.4 Intergranular corrosion test*

Fig. 14a presents the DLEPR curves for the coarse-grained and UFG steel samples following RS. Fig. 14a shows that the passivation curve of the coarse-grained steel exhibits two maxima: the first maximum ($i_{max(1)}$ ~ 30 mA/cm$^2$) is observed at a potential $E_{max(1)}$ ~ 430 mV, and the second maximum ($i_{max(2)}$ ~ 0.1 mA/cm$^2$) at $E_{max(2)}$ ~ 730 mV. In our opinion, the two-stage nature of passivation is associated with the presence of δ-ferrite particles in the austenitic steel (Fig. 2a), which are passivated at high potential values.

The general appearance of the activation and reactivation curves for the UFG steel is similar to that of the CG steel. However, the values of the $E_{max(1)}$ and $E_{max(2)}$ potentials in the UFG steel are shifted to lower values (Fig. 14a). At the same time, the range of potentials $\Delta E_p$ corresponding to the passivation region in the CG steel is $\Delta E_p$ ~ 500 mV and is comparable to the same parameter in the UFG steel ($\Delta E_p$ ~ 440–560 mV). The maximum values of the current density $i_{max(1)}$ in the UFG steel are close to the corresponding values of $i_{max(1)}$ in the CG steel, while the values $i_{max(2)}$ in the UFG steel are slightly less than in the CG steel. On the reactivation curves of the samples cut from the central part of the rod, there is practically no second maximum corresponding to passivation of δ-ferrite particles.

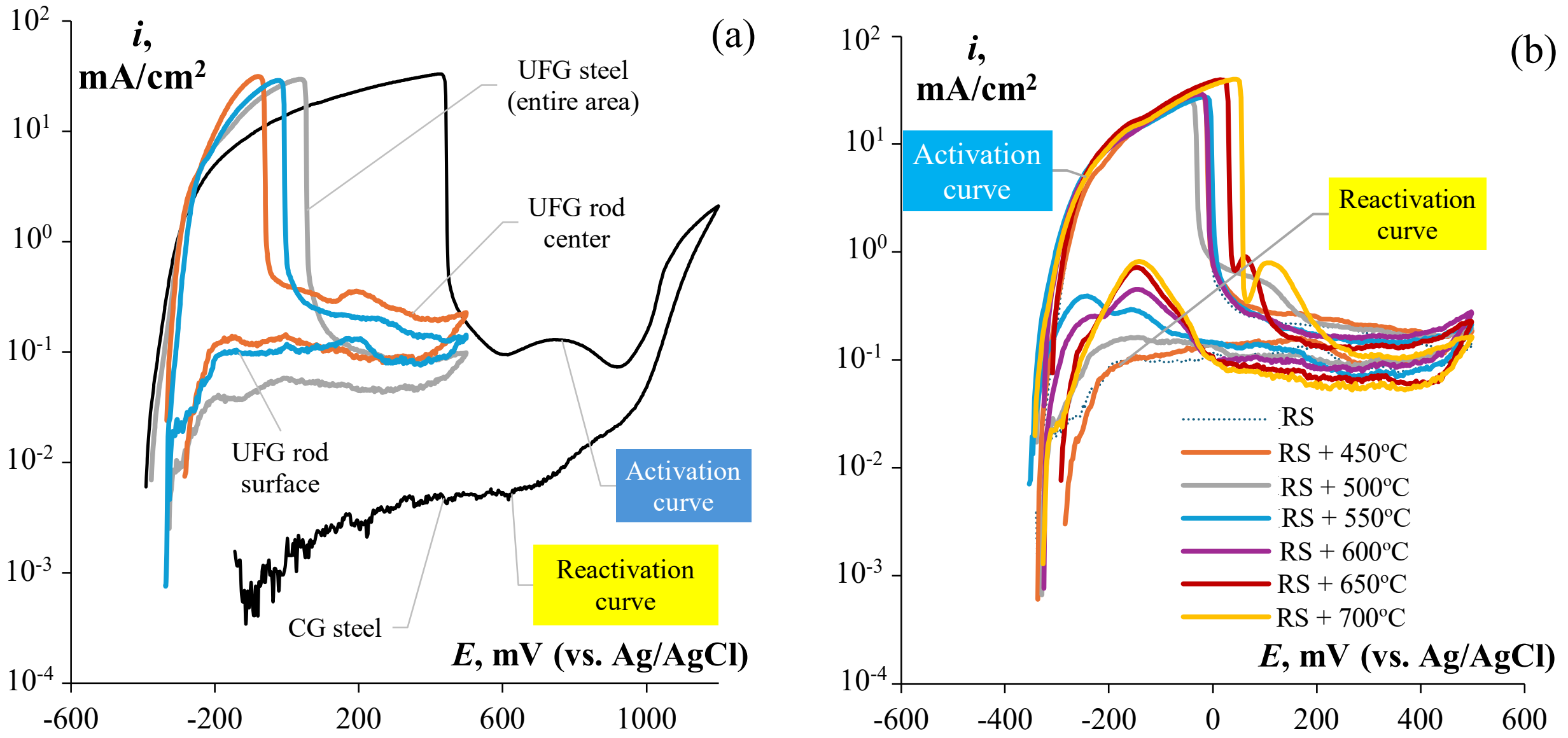

**Fig. 14** DLEPR curves for the 321 steel samples in different states: (a) comparison of CG and UFG steel; (b) influence of annealing temperature (450–700°C) on the appearance of the DLEPR curves of the samples cut from the central part of the UFG steel rod

The formation of a UFG structure by RS leads to a significant decrease in the area under the DLEPR activation curves, while the maximum reactivation current density for the UFG samples is also significantly higher than the reactivation current density for the CG samples (Fig. 14a). This finding indicates that the UFG steel obtained by cold RS exhibits a significantly higher susceptibility to IGC compared to the CG steel in the as-received state: $\langle K_{RS} \rangle = 3.2 \cdot 10^{-3} \gg K_{CG} = (0.1\text{–}0.3) \cdot 10^{-3}$.

The susceptibility of the UFG steel rod's surface layer to IGC (Zone I in Fig. 1c) is several times higher than that of the rod center to IGC (Zone II in Fig. 1c): $K_{RS\text{-}ZoneI} \sim 13.9 \cdot 10^{-3} \gg K_{RS\text{-}ZoneII} \sim 2.7 \cdot 10^{-3}$. The average effective value of the $K$ coefficient determined for the entire area of the UFG samples is $\langle K_{RS} \rangle = 3.2 \cdot 10^{-3}$ (Fig. 1c). The reduced resistance to IGC of the surface layer is probably associated primarily with the increase in the martensite content in Zone I of the steel rod during RS.

It is also interesting to compare the corrosion resistance of the UFG steel 321 samples processed by different severe plastic deformation methods. Previous studies [36] have demonstrated that samples produced by warm ECAP exhibit a lower susceptibility to IGC compared to those processed by cold RS ($K_{ECAP}$ values ranging from $1.64 \cdot 10^{-4}$ to $3.25 \cdot 10^{-4}$ [36]). We attribute the reduced IGC resistance of the cold-RS-processed samples primarily to their higher martensite content.

Fig. 14b illustrates the DLEPR curves for the UFG steel samples that have been annealed at different temperatures. For ease of comparison, the dotted line in Fig. 14b demonstrates the DLEPR curve for the unannealed UFG steel. It can be seen from Fig. 14b that annealing leads to an increase in the area under the reactivation curve and the appearance of a peak corresponding to δ-ferrite on the activation curve.

Annealing at 500–550°C results in a significant increase in the area under the reactivation curve and an increase in the $K$ coefficient (Fig. 15). Consequently, heating to a temperature of 500–550°C is associated with a sharp decrease in the resistance of the UFG steel to IGC. The corrosion resistance of the UFG rod surface after heating to 500°C decreases to a value close to the critical value: $K_{RS\text{-}ZoneI} = 57.1 \cdot 10^{-3} > K_{RS\text{-}ZoneII} = 42.2 \cdot 10^{-3}$. It should be noted that the magnitude of change in the $K$ coefficient during annealing of ECAP-processed UFG samples is much smaller [36]. Therefore, the significant increase in the $K_{RS}$ coefficient to a value close to the critical $K_{IGC}$ is an unexpected and adverse effect that requires further in-depth analysis.

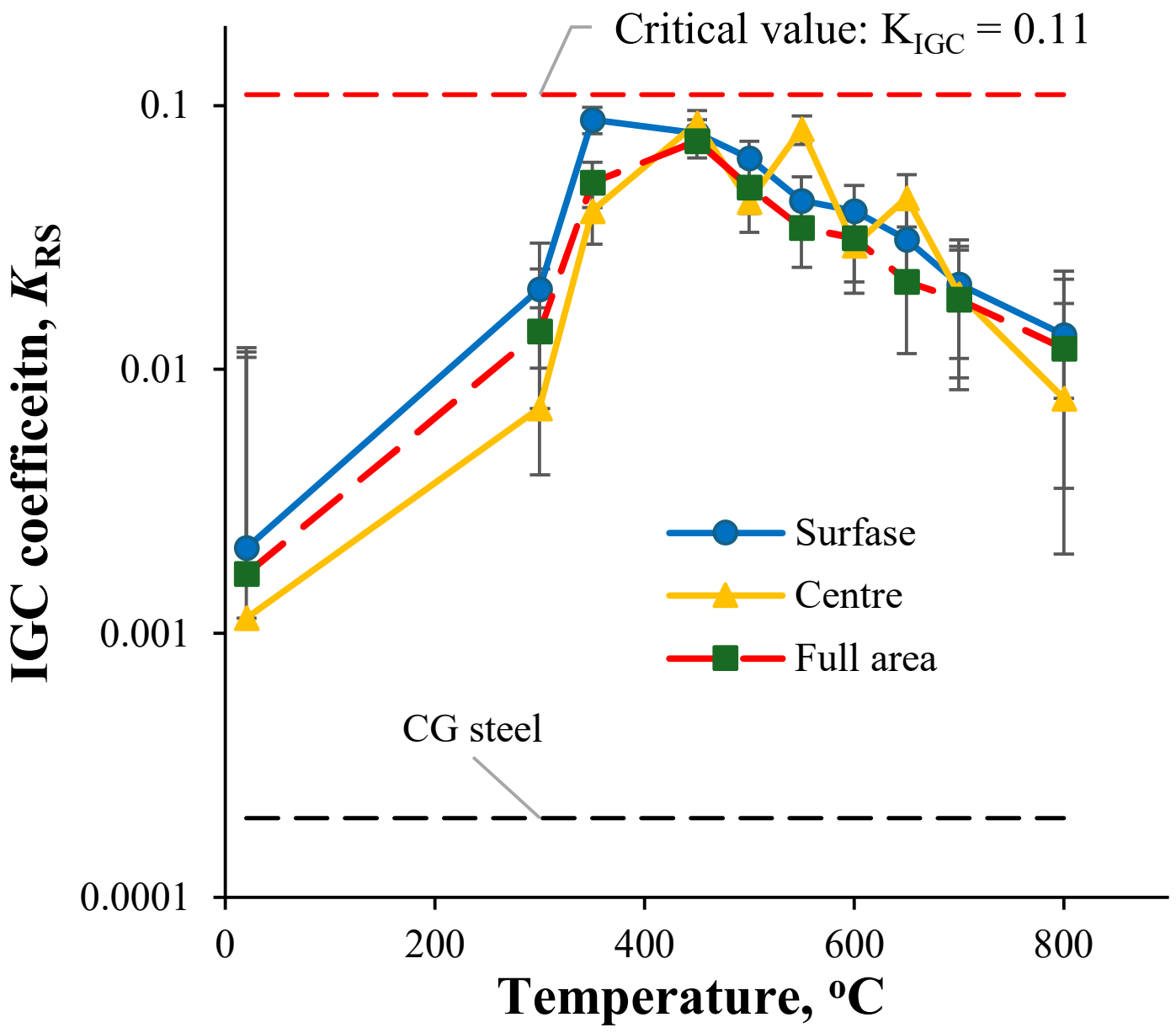

**Fig. 15** Dependence of the $K$ value on the annealing temperature of the UFG steel. The analysis of the effect of the annealing temperature on the IGC susceptibility

A further increase in the annealing temperature leads to a smooth decrease in the IGC rate. After annealing at 800°C, the $K_{RS}$ coefficient decreases to ~(8–12)·$10^{-3}$ (Fig. 15). This indicates that the susceptibility of the annealed UFG steel samples to IGC remains several times higher than that of the coarse-grained steel samples to IGC ($K_{CG}$ = (2 ± 1)·$10^{-3}$).

The increased intensity of corrosion damage in the annealed UFG steels is clearly evident when analyzing the surface of the tested samples. As can be seen from Fig. 16a, traces of elongated corrosion defects are visible on the surface of the unannealed samples; these defects are clearly associated with δ-ferrite particles. On the surface of the samples after corrosion tests, there are scratches from mechanical grinding, which indirectly indicates a low rate of general corrosion of the UFG steel during the DLEPR test. When the annealing temperature is increased to 500–600°C, a sharp increase in the corrosion rate of the sample surfaces is observed (Fig. 16b). After annealing at 700–800°C, the corrosion intensity decreases sharply again, and primarily corrosion damage to δ-ferrite particles is observed (Fig. 16c).

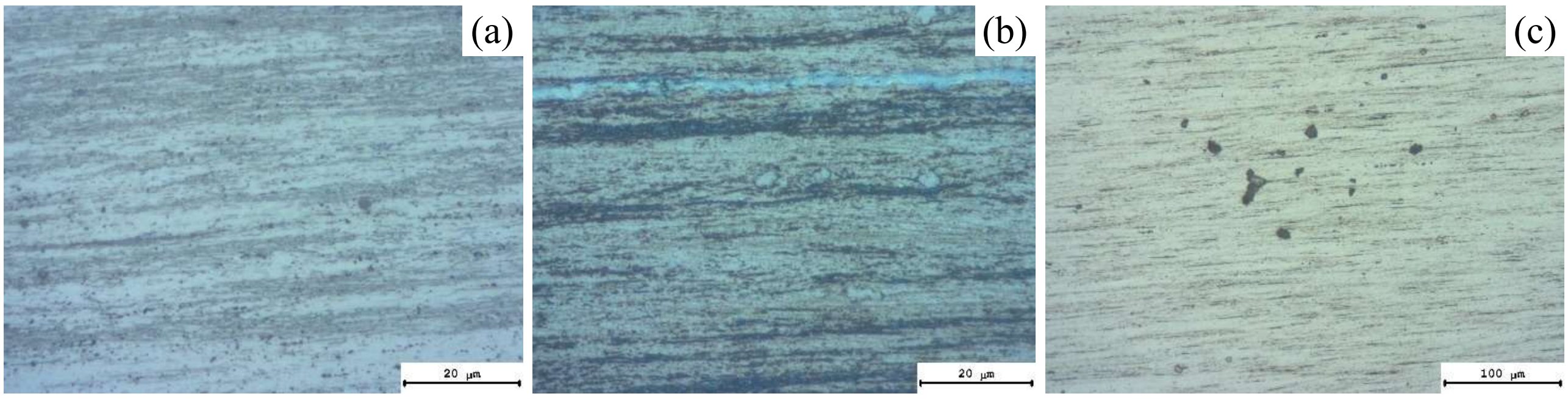

**Fig. 16** OM images of the surface of the UFG samples after DLEPR test: (a) unannealed steel; (b) steel annealed at 550°C; (c) steel annealed at 700°C

Thus, the IGC resistance of the UFG steel is lower than that of the CG steel in its initial state across the entire range of the annealing temperatures. However, it is important to emphasize that the material only approaches the critical IGC susceptibility threshold ($K_{IGC}$ = 0.11) only after annealing at 450–550°C. Notably, at annealing temperatures of 450–500°C, the UFG steel samples exhibit elevated yield strength (YCS, YTS) and ultimate tensile strength (UTS). This result suggests the need for alternative heat treatment methods to simultaneously achieve high strength and IGC resistance in 321 steel samples produced by cold RS. In our opinion, one of the most promising ways to address this issue may be preliminary heat treatment of steel rods using a quenching and high-temperature aging, which eliminates the precipitation of TiC particles during the annealing stage. Another possible option for reducing the susceptibility of resistance to IGC may be rapid heat treatment, which will eliminate the precipitation of TiC particles and form a homogeneous, fine-grained austenitic microstructure devoid of martensite particles.

## 4. Discussion

### *4.1 Structure and mechanical properties*

The mechanisms of UFG structure formation during RS of metastable steels have been extensively studied in recent studies [14, 15, 17, 19–21]. In our view, it is unnecessary to revisit the subject. Therefore, we will limit ourselves to a brief analysis of the results that are important for interpreting corrosion and fatigue tests of the UFG steel.

i) First, it is noteworthy that the as-received coarse-grained 321 steel exhibits an exceptionally high tensile strength (UTS = 680 MPa). Typically, the UTS of austenitic 321 steel (or its equivalent, 08Cr18Ni10Ti) rarely exceed 480–490 MPa, even with a comparable elongation to failure of 35–40%. This enhanced strength in the coarse-grained 321 steel is attributed to the presence of elongated, thin δ-ferrite particles, which effectively impede dislocation slip and propagation of plastic deformation [36]. This observation aligns well with the findings reported in [50, 51].

During the RS process, partial refinement of δ-ferrite particles occurs, which reduces their contribution to the resistance to deformation and fracture.

ii) RS is based on surface plastic deformation of metal workpieces. This leads to the formation of a highly deformed structure with a submicron grain size and an increased content of strain-induced martensite. To increase the content of strain-induced martensite and form an extremely refined austenite structure during subsequent heating, forced cooling of the rod is used during RS. Studies [14, 15, 17, 19-21] reported the possibility of achieving a strain-induced martensite content of 80% in a steel 321 rod after RS with forced cooling. If the steel rod is heated during RS, then there is a decrease in the strain-induced martensite content and a decrease in internal stresses.

iii) The results of the magnetic studies show that after RS, a relatively low content of martensite (~12%, Fig. 5a) is formed in the steel rod without forced cooling, while the average grain size of austenite is refined to a submicron level (Fig. 2d). During the RS process, tensile internal stresses are formed in the steel rod, the value of which is close to the yield strength of the UFG steel. This is probably one of the reasons for the low ductility of the UFG steel in the post-RS state (Table 2).

iv) It was shown in [14, 15, 17, 19–21] that at low strains by RS, the surface hardness of steel rods exceeds the hardness of the central part. At high strains, the nature of the microhardness distribution in the cross-section of the unannealed rod changes: the maximum microhardness values are observed at the surface and in the center of the rod, and in the near-surface region of the steel rod, the Hv values are lower than at the surface and in the center of the rod. A similar result was obtained in our work (Fig. 8a).

v) Annealing at 500°C initiates the reverse transformation of martensite into austenite, which is completed after a 30-minute isothermal hold at 650°C. At the same time, recrystallization begins in the fine-grained austenitic steel. During the initial stages of annealing, this results in a different-grained microstructure, accompanied by a notable reduction in both strength and microhardness. Following a 30-minute hold at 700°C, the sample center exhibits its minimum microhardness and a larger austenite grain size (Fig. 8a). After annealing at 800°C, the samples have a uniform fine-grained austenite microstructure with good ductility.

vi) Upon heating to 550–600°C, the precipitation of nano- and submicron-sized secondary TiC particles begins (Fig. 5). Since this process occurs simultaneously with the reverse transformation of martensite to austenite, its contribution to increasing the microhardness is small and does not significantly affect the strength characteristics of the UFG steel (Table 2).

The approach to analyzing the mechanical properties of the UFG steel 321 was described in [36], and there is no need to dwell on this issue in detail here again. Using the data of [36], it is necessary to highlight only a number of the most important results for further analysis, reflecting the specific features of the UFG structure formed by RS.

A joint analysis of the study results of mechanical properties (Table 2) and structure (Fig. 7) shows that the Hall-Petch relationship is well fulfilled in the steel under study (Fig. 17). The coefficient of determination of the linear approximation of the least squares method is quite high ($R^2$ = 0.92–0.99), indicating good validity of the Hall-Petch relationship. Fig. 17 also shows that the Hall-Petch relationship is well satisfied for all the mechanical properties under study.

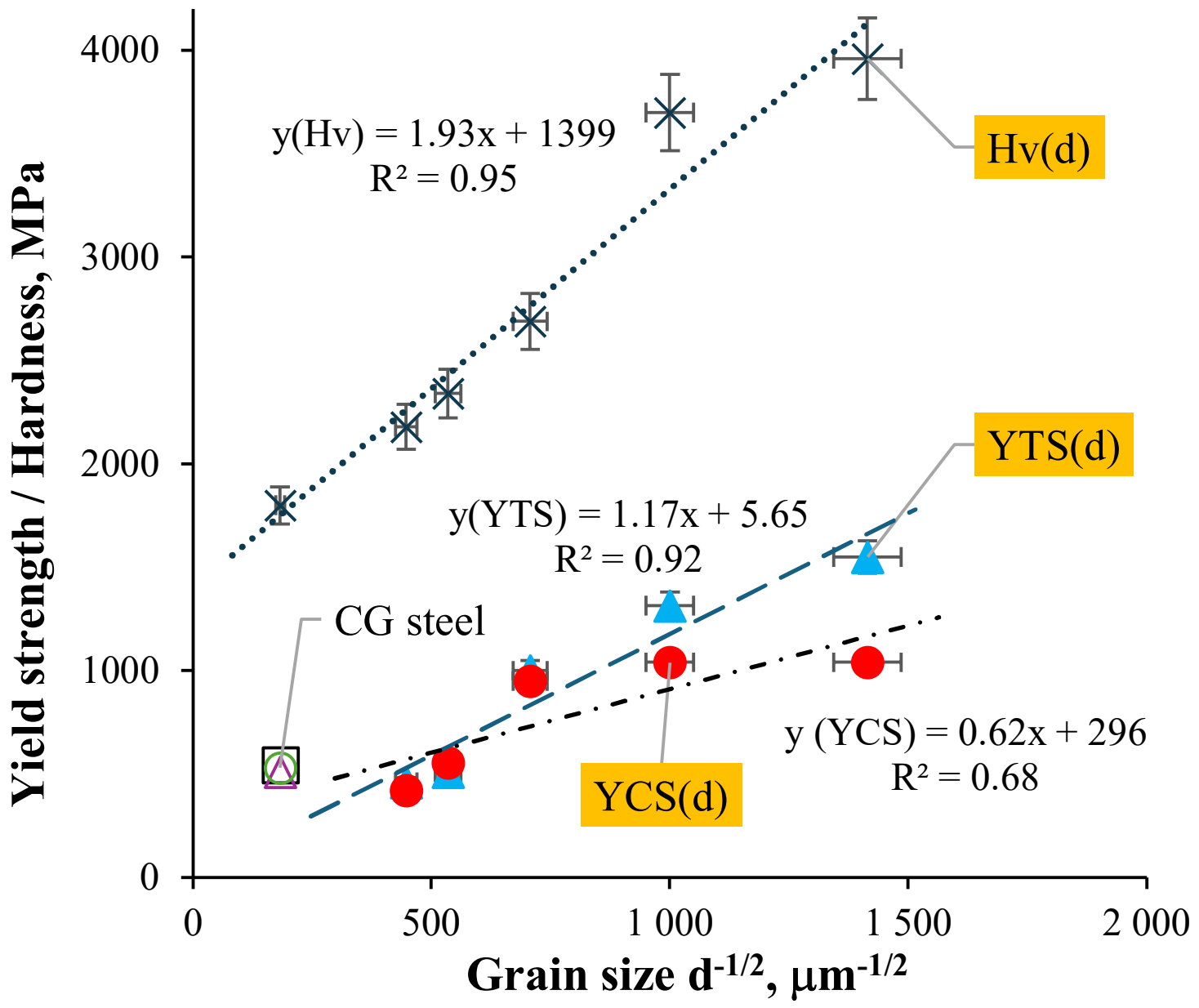

**Fig. 17** Dependences of yield strength (YCS, YTS) and hardness (Hv) on the average grain size of austenite. Verification of the validity of the Hall-Petch relationship in steel 321

*4.2 Corrosion resistance*

The electrochemical studies conducted show that the formation of the UFG microstructure using RS does not have a significant effect on the susceptibility of steel 321 to general corrosion (paragraph 3.2.2). The main mechanism of corrosion damage to the surface of samples during Tafel tests is the formation of corrosion pits near δ-ferrite particles and large Ti(C,N) particles (Fig. 12). Slight differences are observed in the electrochemical characteristics between the center and the surface of the rod. It was noted that increasing the annealing temperature leads to a slight decrease in the corrosion rate by ~0.4 mm/year (Table 3). In general, it can be noted once again that the Tafel method demonstrated low sensitivity to changes in the steel characteristics during RS and annealing.

Steady-state pitting corrosion susceptibility tests have shown that, along with the influence of δ-ferrite particles and large Ti(C,N) inclusions formed during casting, secondary Ti(C,N) nanoparticles have a significant effect. The precipitation of secondary nanoparticles leads to the appearance of small corrosion pits uniformly distributed over the surface of the samples (Fig. 13g) and a non-monotonic change in the relative passivation current $i_{min}/i_0$ (Fig. 13c). It has been established that the increased susceptibility of the annealed steel to pitting corrosion is due to a reduction in the size of δ-ferrite particles. It should be noted that in coarse-grained steels with an equilibrium crystal lattice, this process occurs at higher temperatures [2].

However, in accordance with the general understanding of the features of diffusion processes and grain-boundary phase transformations in highly deformed metals [52–54], it can be assumed that the accelerated dissolution of δ-ferrite particles during annealing of the UFG steel is associated with their partial fragmentation during RS (Fig. 3) and the non-equilibrium state of the crystal lattice and grain boundaries containing an increased density of defects. In particular, non-equilibrium grain boundaries in UFG metals possess a high density of defects, resulting in excess free volume [52]. According to the theory of non-equilibrium grain boundaries, this increased free volume exponentially enhances the grain boundary diffusion coefficient [52]. During low-temperature recrystallization, migrating grain boundaries absorb the high density of lattice dislocations and other structural defects inherent to the highly deformed UFG metal. This leads to a flow of defects entering the grain boundaries, with an intensity proportional to both the boundary migration rate and the lattice dislocation density [52]. According to [52], this leads to an additional increase in the grain boundary diffusion coefficient and the realization of the effect of accelerated diffusion during grain boundary migration in UFG metals. In our opinion, the combination of these factors may ensure an increase in the dissolution rate of finely ground δ-ferrite during the recrystallization annealing of the UFG steel.

The most interesting results are related to the study of the resistance of the UFG steel to IGC by the DLEPR method.

Next, we will discuss the reasons for the unexpected and significant decrease in the corrosion resistance of UFG steel obtained by, as compared to both coarse-grained steel and UFG steel processed by ECAP. It is generally assumed that the primary cause of reduced corrosion resistance in cold-worked austenitic steels is the formation of strain-induced martensite [12, 26–28, 36]. As previously noted, the strain-induced martensite phase forms microgalvanic couples with the surrounding austenite grains, thereby promoting accelerated localized corrosion. A comparison of the present results with the data presented in [36] reveals that UFG steel samples after RS exhibit a higher martensite content than those processed by warm ECAP. This assumption is also supported the comparative analysis of the corrosion resistance at the sample surface (Zone I) versus the sample center (Zone II): $K_{\text{RS-ZoneI}} \sim 13.9\cdot10^{-3} >> K_{\text{RS-ZoneII}} \sim 2.7\cdot10^{-3}$.

As highlighted earlier, annealing at 450–550°C induces an abnormal decrease in the IGC resistance of the UFG steel. Since the low corrosion resistance of the UFG steel is primarily due to the presence of strain-induced martensite, the reverse transformation of martensite into austenite should lead to an increase in the IGC resistance.

Two primary mechanisms are proposed to explain the degradation in corrosion resistance observed during the annealing of UFG steel 321.

First, the reason for the decrease in corrosion resistance is the precipitation of secondary nano- and submicron TiC particles during annealing of the UFG steel. As demonstrated in [55], the free energy of the TiC/γ-Fe interphase boundary is significantly lower than that of the $Cr_{23}C_6$/γ-Fe interphase boundary. Consequently, the precipitation of TiC particles is less dangerous than the precipitation of chromium carbide particles or σ-phase particles. Furthermore, as noted previously, during annealing, these TiC particles precipitate uniformly in the volume of the deformed, without preferential segregation along grain boundaries (Fig. 6). Nevertheless, the precipitation of TiC particles will lead to the formation of additional microgalvanic couples and, as a consequence, to a decrease in the corrosion resistance of the UFG austenitic steel. It is assumed that the more intense precipitation of TiC particles in the surface layer of the steel rod is also one of the factors leading to a decrease in the corrosion resistance of steel in Zone I: $K_{\text{RS-ZoneI}} >> K_{\text{RS-ZoneII}}$.

Notably, the precipitation of TiC particles and the onset of the reverse transformation of martensite into austenite in steel 321 occur in the same temperature range. This suggests that the initiation of the reverse transformation actively promotes TiC precipitation.

30-minute annealing at temperatures exceeding 600–650°C is accompanied by the completion of the reverse transformation of martensite into austenite. This leads to the disappearance of microgalvanic martensite-austenite pairs and an increase in the corrosion resistance of the fine-grained steel. A further slight decrease in the corrosion rate and a decrease in the $K_{RS}$ coefficient is associated with grain boundary migration, which leads to a decrease in the defect density. In accordance with [52], at relatively low grain boundary migration rates, recovery processes are sufficiently intense to induce the near-complete diffusional annihilation of defects at the grain boundaries. This, in turn, reduces both the free energy and the diffusivity of the grain boundaries, thereby diminishing their contribution to the IGC rate.

## Conclusion

1. Cold rotary swaging (RS) of steel 321 reduces the average austenite grain size to the submicron level, increases the strain-induced martensite content, and refines the δ-ferrite particles. Following cold deformation to a true strain of $e$ = 91% (diameter reduction from ∅20 mm to ∅6 mm), the rods of steel 321 exhibit a non-uniform microhardness distribution, with a maximum in the rod center. The UFG steel 321 processed by RS has a reduced resistance to intergranular corrosion (IGC), with an IGC rate significantly higher than that of UFG steel produced by ECAP. Furthermore, the surface layer of the steel rod demonstrates a higher susceptibility to IGC compared to its center. The primary cause of the reduced corrosion resistance in the RS-processed UFG steel is its significantly higher strain-induced martensite content compared to the ECAP-processed UFG steel.

The general corrosion rate and the susceptibility to pitting corrosion remain largely unaffected by the RS processing.

2. 30-minute annealing at 450–550°C leads to a slight increase in microhardness (up to 4.3–4.4 GPa) and ultimate tensile strength (up to 1700–1780 MPa), while simultaneously causing a sharp decrease in IGC resistance. Resistance to pitting corrosion decreases slightly, and the rate of general corrosion does not change within the error of measuring the corrosion current density. The susceptibility of steel to IGC is close to a critical value, remaining significantly higher than that of annealed UFG steel samples produced by ECAP. The primary cause for the decrease in the resistance of UFG steel to localized forms of corrosion is the precipitation of nano- and submicron TiC particles. To reduce the influence of this negative effect, it is recommended to employ a preliminary solution treatment followed by high-temperature aging, which ensures the controlled precipitation of TiC particles within a coarse-grained material. An additional factor contributing to a decrease in IGC resistance is the accumulation of a high density of defects at non-equilibrium grain boundaries, which consequently elevates their free energy.

3. Annealing the UFG steel at temperatures exceeding 650°C enhances its resistance to both IGC and pitting corrosion, while also slightly reducing the general corrosion rate. The improved corrosion resistance is primarily attributed to the completion of the reverse transformation of martensite into austenite and the significant reduction in defect density at migrating non-equilibrium grain boundaries. The decrease in microhardness and strength of the UFG steel at this annealing stage is well-described by the Hall-Petch relationship.

**Funding:** The present study was supported by Ministry of Science and Higher Education of the Russian Federation (grants #FSWR-2026-0039, FSWR-2026-0010).

**CRediT authorship contribution statement: Aleksey Nokhrin**: Formal analysis, Writing–Original draft, Data curation, Visualization, Supervision; **Mikhail Chegurov**: Investigation (corrosion tests, fractography); **Daniil Zotov**: Investigation (microhardness, SEM, EDS, optical microscopy, magnetic phase analysis), **Vladimir Kopylov**: Investigation (Rotary Swaging, annealing); **Anatoliy Sysoev**: Investigation (tensile tests); **Kristina Rubtsova**: Investigation (XRD investigations); **Nikolay Melekhin:** Investigation (compression tests), Project administration, Funding acquisition, Methodology, Formal Analysis, Writing – review & editing, Recourses.

**Declaration of competing interests:** The authors declare that they have no known conflicts of financial interests or personal relationships that could have appeared to influence the work reported in this paper.

**Data and code availability**: Data will be available on the request.